\documentclass[journal]{IEEEtran}
\usepackage{array}
\usepackage{epsfig}
\usepackage{cite}
\usepackage{epstopdf}
\usepackage{amsfonts,amsmath,amsthm,amssymb,balance}
\usepackage{multirow,balance}
\usepackage{bm}
\usepackage[usenames,dvipsnames]{color}
\usepackage{colortbl}
\usepackage{float}
\usepackage[ruled,vlined,linesnumbered]{algorithm2e}
\usepackage{graphicx}
\usepackage{thmtools}
\usepackage{subcaption}
\usepackage{url}
\definecolor{mygray}{gray}{.9}

\graphicspath{{figure/}}
\newcolumntype{C}[1]{>{\PreserveBackslash\centering}p{#1}}
\newcolumntype{R}[1]{>{\PreserveBackslash\raggedleft}p{#1}}
\newcolumntype{L}[1]{>{\PreserveBackslash\raggedright}p{#1}}
\newcommand{\AlgoResetCount}{\renewcommand{\@ResetCounterIfNeeded}{\setcounter{AlgoLine}{0}}}
\newcommand{\AlgoNoResetCount}{\renewcommand{\@ResetCounterIfNeeded}{}}
\newcounter{AlgoSavedLineCount}

\theoremstyle{definition}
\newtheorem{theorem}{Theorem}

\newtheorem{corollary}{Corollary}
\newtheorem{lemma}{Lemma}

\begin{document}
\title{Minimization of Secrecy Outage Probability in Reconfigurable Intelligent Surface-Assisted MIMOME System}

\author{ \IEEEauthorblockN{Yiliang Liu,~\IEEEmembership{Member,~IEEE}, Zhou Su,~\IEEEmembership{Senior Member,~IEEE},  Chi Zhang, and Hsiao-Hwa Chen,~\IEEEmembership{Fellow,~IEEE}}

\thanks{Y. Liu (email: {liuyiliang@xjtu.edu.cn}) and Z. Su (email: {zhousu@ieee.org}) are with the School of Cyber Science and Engineering, Xi'an Jiaotong University, Xi'an, China. C. Zhang (email: {maye1998@163.com}) is with the School of Electronics and Information Engineering, Harbin Institute of Technology, Harbin, China. H.-H. Chen (email: {hshwchen@mail.ncku.edu.tw}) is with the Department of Engineering Science, National Cheng Kung University, Tainan, Taiwan. 
}

}

\maketitle

\begin{abstract}
This article investigates physical layer security (PLS) in reconfigurable intelligent surface (RIS)-assisted multiple-input multiple-output multiple-antenna-eavesdropper (MIMOME) channels. Existing researches ignore the problem that secrecy rate can not be calculated if the eavesdropper's instantaneous channel state information (CSI) is unknown. Furthermore, without the secrecy rate expression, beamforming and phase shifter optimization with the purpose of PLS enhancement is not available. To address these problems, we first give the expression of secrecy outage probability for any beamforming vector and phase shifter matrix as the RIS-assisted PLS metric, which is measured based on the eavesdropper's statistical CSI. Then, with the aid of the expression, we formulate the minimization problem of secrecy outage probability that is solved via alternately optimizing beamforming vectors and phase shift matrices. In the case of single-antenna transmitter or single-antenna legitimate receiver, the proposed alternating optimization (AO) scheme can be simplified to reduce computational complexity. Finally, it is demonstrated that the secrecy outage probability is significantly reduced with the proposed methods compared to current RIS-assisted PLS systems.
\end{abstract}


\section{Introduction}
Serious concerns on information security have been raised in 5G and beyond eras due to the broadcasting nature of wireless channels and hundreds of millions of vulnerable connected devices \cite{Ahmad2019}. Nowadays, physical layer security (PLS) has attracted much attention for strengthening information security as it is capable of achieving confidential information transmission by exploring random characteristics of the wireless medium. The information-theoretic security and cryptography-free characteristic of the PLS technology have received intensive research interests in wireless communications and networks \cite{Wu2019}.

The secrecy capacity of PLS highly depends on the channel difference between legitimate users and eavesdropping users, whereas the channel of legitimate users is usually highly correlated with the eavesdropping channels in most of wireless scenarios, resulting in a low secrecy capacity or secrecy rate \cite{Wu2019}. Although a vast amount of researches report that multiple antennas, cooperative relay, and artificial noise technologies can improve secrecy capacity, the deployment of numerous radio frequency modules incurs excessive hardware cost \cite{Liusurvey2017}. Moreover, artificial noise signals require extra transmission power for security guarantees. Recently, RIS has attracted wide research attention from both industry and academia because of its abundant spatial degrees of freedom and controllability. It is composed of a large number of low-cost passive reflecting elements that can control the direction of the electromagnetic wave by adaptively adjusting the phase shifter of each element, and inhibits the signal from being leaked at a certain position  \cite{Wuirs2019}, whereby the channel difference between legitimate users and eavesdropping users is enlarged, and the secrecy capacities of legitimate users can be improved consequently. Following that, plenty of researches have been conducted to investigate RIS-assisted PLS, including information-theoretic aspects \cite{Yang2020,Trigui2021,hong2020outage, Zhangtifs2021, hou2019mimo} and secure beamforming with phase shifter adjustments \cite{Miao2019, Hong2019, Qiao2020, Zheng2021,HongAN2020,HongRobust2021,Dong2020,Shu2021,Niu2021,  Yu2020,Elhoushy2021, Feng2020}. 

The instantaneous CSIs of eavesdroppers are usually unknown when the eavesdroppers keep silent, which is a usual issue but ignored in recent RIS-assisted PLS researches \cite{Niu2021,Qiao2020,Hong2019,Dong2020,Miao2019,Yu2020, Shu2021,Zheng2021, HongRobust2021, Elhoushy2021,HongAN2020}. The above works maximize the secrecy capacity or secrecy rate via beamforming and phase shifter optimization, assuming that instantaneous CSIs of eavesdroppers are available. However, the secrecy capacity or secrecy rate can not be measured if instantaneous CSIs of eavesdroppers are unknown. To address the eavesdropper's CSI issue, typical methodologies are secure beamforming technologies with secrecy outage-based PLS coding \cite{Harrison2013,LiuICV2021}, where secure beamforming can reduce the secrecy outage probability. Regarding the secrecy outage probability as a threshold, the PLS coding rate is adjusted during confidential data transmission. However, the existing expressions of secrecy outage probability for RIS-assisted PLS focus on single-antenna scenarios \cite{Yang2020, Trigui2021, hong2020outage} or are deduced with fixed phase shifter matrices and beamforming vectors \cite{Zhangtifs2021}, which are not suitable for beamforming and phase shifter optimization. Feng \textsl{et al.} proposed an alternating guideline to solve this eavesdropper's CSI problem \cite{Feng2020}. They simulate a great deal of the instantaneous eavesdropper's CSIs as samples to calculate an approximate secrecy rate as the optimization objective of beamforming and phase shifter optimization algorithms. The generation process of many CSI samples causes a significant latency, which inevitably increases the cost of computing modules.

According to the above discussion, the first concern in this paper is finding an expression of secrecy outage probability in the MIMOME scenario as the performance metric of RIS-assisted PLS. Then, an optimization algorithm should be designed to minimize the secrecy outage probability. The main contributions of this work are summarized as follows.
\begin{enumerate}
\item Firstly, we formulate the PLS model of a RIS-assisted MIMOME channel, where beamforming and phase shifter are optimized for security purposes. As the eavesdropper's instantaneous CSI is unknown, the optimization objective is to minimize the secrecy outage probability measured with the eavesdropper's statistical CSI.  
\item Due to the mathematical complexity of the RIS-assisted MIMOME channel model, it is hard to get the exact expressions of secrecy outage probability. Here, we use the Gamma distribution to fit the closed-form expression of secrecy outage probability for any beamforming vector and phase shifter matrix.
\item With the expression of secrecy outage probability, we transform the minimization problem of secrecy outage probability into two sub-problems, i.e., beamforming vector and phase shifter matrix optimization, which are solved optimally by generalized Rayleigh quotient and quadratic optimization methods, respectively. Then, an alternating optimization (AO) algorithm is proposed to find the global results for the beamforming vector and phase shifter matrix. Lastly, the proposed scheme is simplified in the single-antenna transmitter or single-antenna legitimate receiver case, which has lower computational complexity than that in the MIMOME channel case.
\end{enumerate}

The remainder of the paper is organized as follows. Section \ref{surveys} surveys the related works. Section \ref{model} describes the system model and problem formulation. The expressions of secrecy outage probability are given in Section \ref{expression}. The optimization algorithms of beamforming vector and phase shifter matrix are proposed in Section \ref{proposed1}. The optimization algorithms in single-antenna cases are presented in Section \ref{scase}. We show simulation results in Section \ref{simulations}, and conclude this paper in Section \ref{conclusions}.

\textsl{Notations:} Bold uppercase letters, such as $\mathbf{A}$, denote matrices, and bold lowercase letters, such as $\mathbf{a}$, denote column vectors. $\mathbf{A}^{\dagger}$, $\mathbf{A}^{\rm{T}}$, and $\mathbf{A}^{\rm{H}}$ represent the conjugate transformation, transpose, and conjugate transpose of $\mathbf{A}$, respectively. $\mathbf{I}_a$ is an identity matrix with its rank $a$. $\mathcal{CN}(\mu,\sigma^2)$ is a complex normal (Gaussian) distribution with mean $\mu$ and variance $\sigma^2$. $(\mathbf{A})^{-1}$ is the inverse function of $\mathbf{A}$. $|\mathbf{x}|$ is the Euclidean norm of $\mathbf{x}$. $\text{diag}(\mathbf{x})$ is the diagonal matrix of $\mathbf{x}$. $\mathbb{E}(\cdot)$ is the expectation operation. $\text{vec}(\mathbf{A})$ is the vectorization of the diagonal elements of $\mathbf{A}$. $\text{arg}(x)$ is the angle of a complex variable $x$. An $a\times(b+c)$ matrix $[\mathbf{A},\mathbf{B}]$ denotes a combined matrix between an $(a\times b)$ matrix $\mathbf{A}$ and an $(a\times c)$ matrix $\mathbf{B}$. $\circ$ is the Hadamard product. $\mathfrak{R}(x)$ means the real part of a complex variable $x$.

\section{Related Works}\label{surveys}

\subsection{Information-Theoretic Research on RIS-assisted PLS}
As mentioned in \cite{Zhangtifs2021}, the information-theoretic research about RIS-assisted PLS is fundamental that provides optimization objectives, constraint functions, as well as performance metrics for beamforming and phase shifter. The channel models with RIS are the main mathematical challenges of this research. Yang \textsl{et al.} considered the RIS-assisted single-antenna case, i.e., every transmitter, legitimate receiver, and eavesdropper has one antenna, then used the Gaussian distribution to get the expression of secrecy outage probability when the instantaneous CSIs of eavesdroppers are unknown \cite{Yang2020}. Furthermore, Trigui \textsl{et al.} gave the expressions of ergodic secrecy rate and secrecy outage probability in the RIS-assisted single-antenna case, simultaneously \cite{Trigui2021}. They used Fox's H transform theory and the Mellin-Barnes integrals to get these expressions. Hong. \textsl{et al.} derived the expression of secrecy outage probability for the RIS-assisted multiple-input single-output multiple-antenna-eavesdropper (MISOME) scenario, where the secrecy outage probability is caused by channel estimation errors \cite{hong2020outage}. The RIS-assisted MIMOME channel model was taken into account by Zhang \textsl{et al.} \cite{Zhangtifs2021}, where the model is based on the zero-forcing phase shifter matrix and maximal ratio combining receiving (MRC) vector proposed in \cite{hou2019mimo} for security purposes. Also, they deduced the expression of ergodic secrecy rate and secrecy outage probability by the $\kappa-\mu$ distribution \cite{Bhargav2018}.

\subsection{Secure beamforming with Phase Shifter Adjustments}
The feature of secure beamforming in RIS-assisted PLS is the phase shifter adjustment. The secrecy rate maximization with respect to the variables of beamforming vectors and phase shifter matrices is a non-convex problem and is usually solved by alternating optimization over individual beamforming and phase shifter \cite{Miao2019, Hong2019, Qiao2020, Feng2020 ,Zheng2021, HongAN2020 , Dong2020, HongRobust2021, Niu2021, Shu2021, Elhoushy2021,Yu2020}. Especially, Cui \textsl{et al.} considered the RIS-assisted MISOME model. In this model, the secrecy rate is maximized by alternating optimization between beamforming vector and phase shifter matrix, in which the subproblems of beamforming and phase shifter optimization are solved by generalized Rayleigh quotient and semidefinite relaxation (SDR), respectively \cite{Miao2019}. Shen \textsl{et al.} simplified the phase shifter optimization via the minorization-maximization (MM) \cite{Hong2019} and element-wise block coordinate descent (BCD)-based methods \cite{Qiao2020}.  Furthermore, Chu \textsl{et al.} introduced the artificial noise (AN) technique in RIS-assisted PLS to improve the secrecy rate by optimizing  transmitting and AN covariance matrices \cite{HongAN2020, Zheng2021}. Then, the works \cite{HongAN2020, Zheng2021} were extended into the scenario with multiple non-cooperative Eves and considered channel estimation errors \cite{HongRobust2021}.  Dong \textsl{et al.} considered the RIS-assisted MIMOME model, and used the MM algorithm to optimize the phase shifter matrix for maximizing the secrecy rate \cite{Dong2020}. Shu \textsl{et al.} improved the secrecy rate of the RIS-assisted MIMOME model via designing closed-form AN precoding, by which AN signals only interfere with Eve but have no effect on the legitimate user \cite{Shu2021}. Niu  \textsl{et al.} formulated the weighted sum secrecy rate maximization problem in the RIS-assisted multiple user MISOME scenarios, where a successive convex approximation (SCA) technique is employed to find the optimal beamforming and AN signals \cite{Niu2021}. Yu \textsl{et al.} extended the work of \cite{Niu2021} by adding the multiple RIS devices and considered the errors of channel estimation \cite{Yu2020}. For the RIS-assisted massive multiple-input multiple-output (MIMO) scenario with an eavesdropper, it is time-consuming to find optimal beamforming and phase shifter by alternating optimization because the dimensionality of parameters is very large. In this case, Elhoushy \textsl{et al.} simplified the RIS-assisted PLS scheme by using closed-form beamforming, i.e.,  the zero-forcing (ZF) precoding technology, then found the optimal  phase shifter matrix to improve the secrecy rate~\cite{Elhoushy2021}. Nowadays, the RIS device is deployed in the millimeter-wave system where a stochastic learning method is presented to tackle the random blockage problem~\cite{Zhou2021}. 

\subsection{Discussion}
The existing investigations of RIS-assisted PLS do not concern secrecy outage probability in the MIMOME model. Besides, the optimization algorithm should be presented for secrecy outage probability minimization. In this work, we first provide the expression of secrecy outage probability in the RIS-assisted MIMOME channel for any beamforming vector and phase shifter matrix, followed by secure beamforming and phase shifter optimization schemes that can reduce the secrecy outage probability.

\section{System Model and Problem Formulation}\label{model}

Considering that the eavesdropper's instantaneous CSI is unknown for legitimate users, the channel model of RIS-assisted MIMOME is established in this section. 

\subsection{Channel Model}

\begin{figure}[t]
\centering
\includegraphics[width=0.9\linewidth]{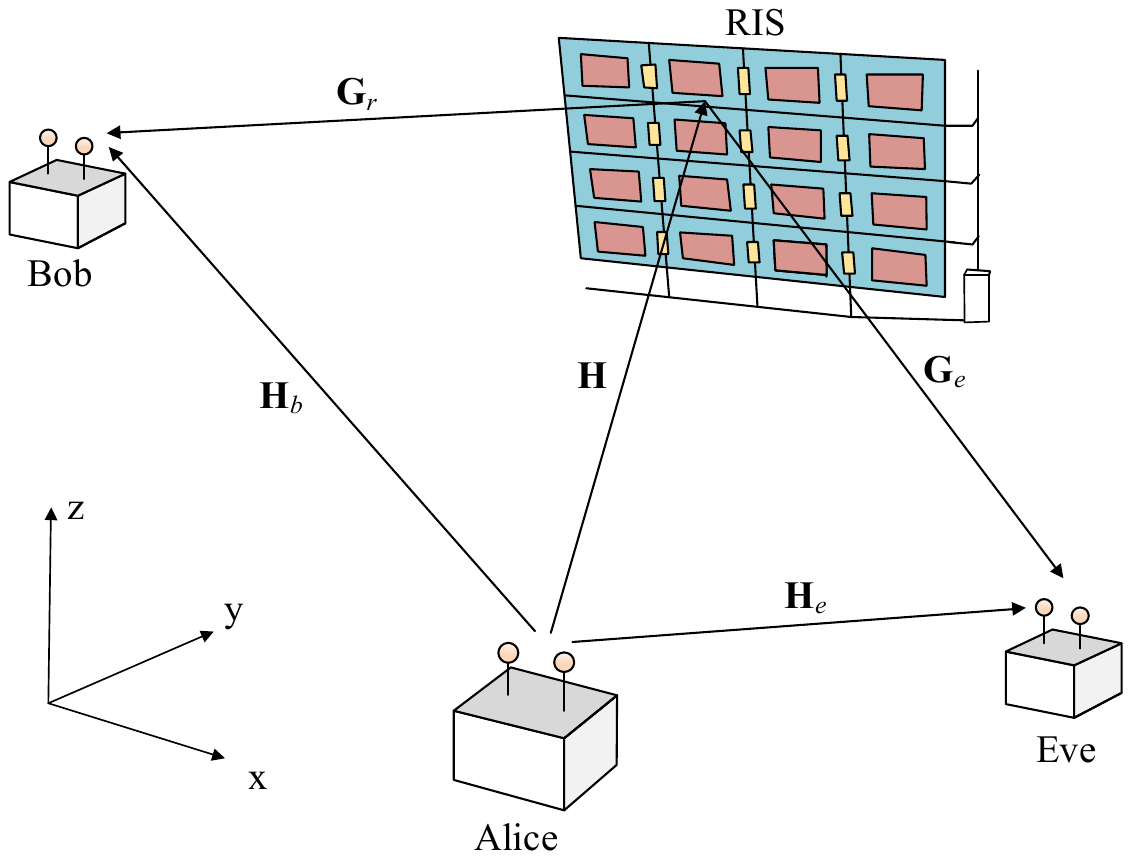}
\caption{RIS-assisted MIMOME model, where $(\alpha \mathbf{H}_b+\mathbf{G}_r\bm{\Phi}\mathbf{H})$ and $(\beta \mathbf{H}_e+\mathbf{G}_e\bm{\Phi}\mathbf{H})$ are defined as the main channel and wiretap channel, respectively. }\label{model_figure}
\end{figure}

This article considers a RIS-assisted MIMOME channel, as shown in Fig. \ref{model_figure}, including a transmitter (Alice) with $N_t$ antennas, an eavesdropper (Eve) with $N_e$ antennas, a RIS equipped with $N_s$ programmable phase shifter elements, and a legitimate user (Bob) with $N_r$ antennas. Assume that wiretap channels obey  Rayleigh fading, i.e., the channel from Alice to Eve is defined as $\mathbf{H}_e\sim \mathcal{CN}_{N_e,N_t}(\mathbf{0},\mathbf{I}_{N_e}\otimes \mathbf{I}_{N_t})$, and the channel from RIS to Eve is defined as $\mathbf{G}_e\sim \mathcal{CN}_{N_e,N_s}(\mathbf{0},\mathbf{I}_{N_e}\otimes \mathbf{I}_{N_s})$. The legitimate channels are assumed to be the quasi-static flat-fading model, i.e., the channel from Alice to Bob is defined as $\mathbf{H}_b\in \mathbb{C}^{N_r\times N_t}$, the channel from Alice to RIS is defined as $\mathbf{H}\in \mathbb{C}^{N_s\times N_t}$, the channel from RIS to Bob is defined as $\mathbf{G}_{r}\in \mathbb{C}^{N_r\times N_s}$. The channel estimation for $\mathbf{H}_b$, $\mathbf{H}$, and $\mathbf{G}_r$ is perfect. Let $\alpha \in [0,1]$ and $\beta\in [0,1]$ denote the existence probabilities of the channel from Alice to Bob and the channel from Alice to Eve, e.g., $\alpha =1$ means that there is the direct channel between Alice to Bob, and $\beta =0.3$ means a $30\%$ probability that the direct channel from Alice to Eve is existed. $\alpha$ and $\beta$ are statistical values that can be usually obtained  via historical data. If $\alpha$ and $\beta$ are unknown, we consider the pessimistic scenario $\{\alpha=0,\beta=1\}$ for the security guarantee.

Alice uses beamforming vector $\mathbf{w}\in \mathbb{C}^{N_t\times 1}$ to transmit confidential information-bearing signal $x$ to Bob, where $\mathbf{w}^{\rm{H}}\mathbf{w}=P \leq \rho$, $P$ is the actual transmission power, $\rho$ is the transmission power constraint, and $\mathbb{E}(xx^{\rm{H}})=1$. In addition, RIS controls programmable phase shifter elements via a phase shifter matrix, where the phase shifter matrix is defined as an $N_s\times N_s$ matrix $\bm{\Phi}$, i.e., 
\begin{flalign}
\bm{\Phi}=\text{diag}[\exp(j\theta_1),...,\exp(j\theta_{n}),..., \exp(j\theta_{N_s})],
\end{flalign}
and $\theta_n\in [0,2\pi)$ is the phase introduced by the $n$th phase shifter element of RIS.

With beamforming vector $\mathbf{w}$ and phase shifter matrix $\bm{\Phi}$, the received signals at Bob and Eve can be expressed as
\begin{flalign}
& \mathbf{y} = (\alpha \mathbf{H}_b+\mathbf{G}_r\bm{\Phi}\mathbf{H})\mathbf{w}x+\mathbf{n}, \label{mchannel}\\
& \mathbf{y}_e =(\beta \mathbf{H}_e+\mathbf{G}_e\bm{\Phi}\mathbf{H})\mathbf{w}x+\mathbf{n}_e, \label{Le2}
\end{flalign}
where $(\alpha \mathbf{H}_b+\mathbf{G}_r\bm{\Phi}\mathbf{H})$ and $(\beta \mathbf{H}_e+\mathbf{G}_e\bm{\Phi}\mathbf{H})$ are known as the main channel and wiretap channel, respectively. $\mathbf{n}$ and $\mathbf{n}_e$ is the additive white Gaussian noise (AWGN) obeying $\mathcal{CN}_{N_r,1}(\mathbf{0},\sigma^2\mathbf{I}_{N_r})$ and $\mathcal{CN}_{N_e,1}(\mathbf{0},\sigma_e^2\mathbf{I}_{N_e})$, respectively. 

Considering a pessimistic condition that Eve knows the instantaneous CSI of all channels, including $\mathbf{G}_r$, $\mathbf{H}_b$, $\mathbf{H}$, $\mathbf{G}_e$, $\bm{\Phi}$, and $\mathbf{H}_e$, while Alice, Bob, and RIS only know the instantaneous CSIs of legitimate devices, including $\mathbf{G}_r$, $\mathbf{H}_b$, $\mathbf{H}$, and $\bm{\Phi}$. Since instantaneous CSIs $\mathbf{H}_e$ and $\mathbf{G}_e$ are unknown for legitimate users, the straightforward PLS schemes try to improve the channel capacity at Bob \cite{Liusurvey2017}. With the maximal-ratio transmission (MRT) technology of MIMO \cite{Kang2003}, the phase shifter matrix is usually adjusted for maximizing diversity gain, i.e., $\max_{\bm{\Phi}}\lambda_{\max}[(\alpha\mathbf{H}_b+\mathbf{G}_{r}\bm{\Phi}\mathbf{H})^{\rm{H}}(\alpha\mathbf{H}_b+\mathbf{G}_{r}\bm{\Phi}\mathbf{H})]$, where $\lambda_{\max}(\mathbf{A})$ is the largest eigenvalue of $\mathbf{A}$. It is just a method of channel capacity improvement but is not a desired design for secrecy performance of RIS-assisted PLS. Our paper is to find the optimal $\mathbf{w}$ and $\bm{\Phi}$ to minimize the secrecy outage probability.

\subsection{Secrecy Rate and Secrecy Outage Probability}

The usual secrecy performance in PLS is secrecy rate \cite{Liusurvey2017}, which is formulated as follows,
\begin{flalign}
C_{s}=(C_{m}-C_{w})^{+}, 
\end{flalign}
where $C_{m}$ and $C_w$ are defined as the main channel capacity and wiretap channel capacity, i.e., 
\begin{flalign}
&C_{m}=\log_2\bigg[1+\frac{1}{\sigma^2}| (\alpha\mathbf{H}_b+\mathbf{G}_r\bm{\Phi}\mathbf{H})\mathbf{w}|^2\bigg], \label{cm}  \\
&C_{w}=\log_2\bigg[1+\frac{1}{\sigma^2_e}|(\beta \mathbf{H}_e+\mathbf{G}_e\bm{\Phi}\mathbf{H})\mathbf{w}|^2 \bigg].
\end{flalign}
Note that $C_m$ is achievable at Bob when using the matching receiving vector, i.e., $\mathbf{w}_r=[(\alpha \mathbf{H}_b+\mathbf{G}_r\bm{\Phi}\mathbf{H})\mathbf{w}]^{\rm{H}}/|(\alpha \mathbf{H}_b+\mathbf{G}_r\bm{\Phi}\mathbf{H})\mathbf{w}|$. Similarly, $C_w$ is achievable at Eve via its receiving vector $\mathbf{w}_e= [(\beta \mathbf{H}_e+\mathbf{G}_e\bm{\Phi}\mathbf{H})\mathbf{w}]^{\rm{H}}/|(\beta \mathbf{H}_e+\mathbf{G}_e\bm{\Phi}\mathbf{H})\mathbf{w}|$. However, due to the lack of $\mathbf{H}_e$ and $\mathbf{G}_e$, it is hard to check whether an instantaneous secrecy rate is nonnegative or not. In this case, PLS usually uses secrecy outage probability as the security metric for PLS coding or optimization algorithms. The secrecy outage probability is defined as the probability that the targeted PLS coding rate of Alice's encoder, i.e., $R_s$, is larger than the secrecy rate $C_s$. From \cite[Eq. (4)]{Zhou2011RethinkingSecrecyOutage}, the secrecy outage probability can be expressed  as 
\begin{flalign}\label{uso}
P_{\text{out}}&=P(C_{s} \leq R_s \big| \text{Transmission})\notag\\
& =P(C_{w}\geq C_{m}-R_s).
\end{flalign}
It is evident that the secrecy outage probability is the conditional probability based on reliable transmissions of main channels, i.e., Bob decodes transmitted codewords correctly with the rate up to $C_{m}$. As we assume that Alice has perfect knowledge about the instantaneous CSIs of $\mathbf{H}_b$, $\mathbf{G}_r$, $\mathbf{H}$, and $\bm{\Phi}$ within coherence time, Alice can use an adaptive rate of transmitted codewords that equals to $C_{m}$.

\subsection{Problem Formulation}
In this article, the optimization problem is mathematically expressed as the minimization of the secrecy outage probability via phase shifter matrix and beamforming vector optimization,
\begin{flalign}
\text{P1: } &\min_{\bm{\Phi},\mathbf{w}}P_{\text{out}},  \\
&\text{s.t. }  |\exp(j\theta_n)|^2=1, \quad n=1,...,N_s, \label{P1C1}  \\ 
& \quad \; \; \mathbf{w}^{\rm{H}}\mathbf{w} \leq \rho, \label{P1C2} 
\end{flalign}
where $P_{\text{out}}$ is the secrecy outage probability, $\mathbf{w}$ is the beamforming vector, and $\bm{\Phi}$ is the phase shifter matrix. Eq. (\ref{P1C1}) corresponds to the unit-modulus requirements of the reflection elements at the RIS. Eq. (\ref{P1C2}) is the transmission power constraint of beamforming vectors. Obviously, the priority of solving this problem is to find the expression of $P_{\text{out}}$.

\section{Expression of Secrecy Outage Probability}\label{expression}
This paper uses Gamma distributions to fit the expression of secrecy outage probability.

\subsection{Preliminaries}
The gain of wiretap channels of the presented system model is seen as a random variable defined by $X$, and we derive the CDF of $X$ in Lemma 1 via Gamma distributions. The CDF of $X$ is used to deduce the expression of secrecy outage probability. The Gamma distribution can represent the sum of multiple independent random variables that obey the exponential distribution, so it is suitable for representing the gain of RIS-assisted channels, such as \cite{VanChien2021, Cui2021,Salhab2021}. 

\begin{lemma}
For any $n\times 1$ complex vector $\mathbf{u}$, $\beta\in [0,1]$, two independent random variables $\mathbf{a}\sim\mathcal{CN}_{m,1}(\mathbf{0},\mathbf{I}_{m})$, and $\mathbf{C}\sim\mathcal{CN}_{m,n}(\mathbf{0},\mathbf{I}_{m}\otimes\mathbf{I}_{n})$, if we have random variable $x$ as
\begin{flalign}\label{x1}
&x=|\beta \mathbf{a}+\mathbf{Cu}|^2,
\end{flalign}
the CDF of $X \sim X(\beta,m,n,\mathbf{u})$ can be expressed as
\begin{flalign}\label{cdfw1}
F_{X}(x)&=1-\frac{1}{\Gamma(m)}\Gamma\bigg(m,\frac{x}{\beta^2+|\mathbf{u}|^2}\bigg),
\end{flalign}
where $\Gamma(x)$ is the Gamma function with respect to $x$, and $\Gamma(\epsilon, \eta)$ is the upper incomplete Gamma function defined as follows,
\begin{flalign}\label{ligamma}
\Gamma(\epsilon,\eta)=\int_{\eta}^{\infty}\exp(-z)z^{\epsilon-1}\text{d}z,
\end{flalign}
\end{lemma}
\begin{IEEEproof}
See in Appendix A. 
\end{IEEEproof}

\subsection{Expression of Secrecy Outage Probability}
Here, we use Lemma 1 to deduce the expression of secrecy outage probability. 

\begin{theorem}[Expression of secrecy outage probability]
The secrecy outage probability\footnote{For the expression of the Rician model, please refer to
\url{https://github.com/yiliangliu1990/liugit_pub/tree/master/IRS}.} of $R_s$, i.e., $P_{\text{out}}$, is expressed as
\begin{flalign}\label{esop}
P_{\text{out}}&=1-F_{X}(\phi_1)=\frac{1}{\Gamma(N_e)}\Gamma\bigg(N_e,\frac{\phi_1}{\beta^2+|\bm{\Phi}\mathbf{Hb}|^2}\bigg),
\end{flalign}
where $\phi_1=\sigma_e^2(2^{C_{m}-R_s}-1)/P$, $C_m$ is defined in Eq. (\ref{cm}), $F_{X}(x)$ is defined in Eq. (\ref{cdfw1}), $|(\beta\mathbf{H}_e+\mathbf{G}_e\bm{\Phi}\mathbf{H})\mathbf{b}|^2\sim X(\beta, N_e,N_s,\bm{\Phi}\mathbf{Hb})$ represents a random variable, $\mathbf{b}=\mathbf{w}/\sqrt{P}$, and $P$ is the actual transmission power with $P\leq\rho$.
\end{theorem}
\begin{IEEEproof}
As channel capacity $C_{m}$ can be calculated by Eq. (\ref{cm}), with a pre-defined $R_s$, we can transform $P_{\text{out}}$ into
\begin{flalign}\label{probability}
P_{\text{out}}&=P(C_{w}>C_{m}-R_s|\text{Transmission})\notag\\
&=P(|(\beta\mathbf{H}_e+\mathbf{G}_e\bm{\Phi}\mathbf{H})\mathbf{b}|^2 \geq\phi_1).
\end{flalign}
Since random matrix $\mathbf{H}_{e}$ is a cyclic symmetry complex Gaussian matrix and fixed $\mathbf{b}$ is an unitary vector, $\mathbf{H}_{e}\mathbf{b}$ is a cyclic symmetry complex Gaussian vector, i.e., $\mathbf{H}_{e}\mathbf{b}\sim\mathcal{CN}_{N_e,1}(\mathbf{0},\mathbf{I}_{N_e})$ \cite{Liu2017SecrecyCapacityAnalysis}. It means that $|(\beta\mathbf{H}_e+\mathbf{G}_e\bm{\Phi}\mathbf{H})\mathbf{b}|^2$ belongs to $X(\beta, N_e, N_s,\mathbf{u})$ in Lemma 1 where $\mathbf{u}=\bm{\Phi}\mathbf{H}\mathbf{b}$. According to Eq. (\ref{cdfw1}), we get the expression of secrecy outage probability as shown in Eq. (\ref{esop}).
\end{IEEEproof}

\begin{corollary}[High SNR case]
In the case of $\frac{P}{\sigma_e^2}\to \infty$, the lower bound of secrecy outage probability can be expressed as 
\begin{flalign}\label{ae}
P_{\text{out}}\geq\frac{1}{\Gamma(N_e)}\Gamma\bigg(N_e,\frac{\sigma_e^2|(\alpha\mathbf{H}_b+\mathbf{G}_r\bm{\Phi}\mathbf{H})\mathbf{b}|^2}{\sigma^22^{R_s}(\beta^2+|\bm{\Phi}\mathbf{Hb}|^2)}\bigg).
\end{flalign}
It means that in high SNR cases, the secrecy outage probability reaches to a constraint, i.e., it decreases with an increasing $P$ at the beginning, but boundlessly increasing the transmission power can not provide the infinite help for reducing secrecy outage probability.
\end{corollary}

\begin{IEEEproof}
In the case of $\frac{P}{\sigma_e^2}\to \infty$, $\phi_1$ can be re-written as
\begin{flalign}\label{hsnrphi}
\phi_1&= \frac{\sigma_e^2(2^{C_{m}-R_s}-1)}{P} \notag \\
&=\frac{\sigma_e^2|(\alpha\mathbf{H}_b+\mathbf{G}_r\bm{\Phi}\mathbf{H})\mathbf{b}|^2}{\sigma^22^{R_s}}+\frac{\sigma_e^2}{P}\bigg(\frac{1}{2^{R_s}}-1\bigg) \notag \\
&\leq \frac{\sigma_e^2|(\alpha\mathbf{H}_b+\mathbf{G}_r\bm{\Phi}\mathbf{H})\mathbf{b}|^2}{\sigma^22^{R_s}}.
\end{flalign}
Since $N_e>0$ and $z=\frac{\phi_1}{\beta^2+|\bm{\Phi}\mathbf{Hb}|^2}>0$, we have $\frac{\partial \Gamma(N_e,z)}{\partial z}=-z^{N_e-1}\exp(-z)$, and it is concluded that $P_{\text{out}}$ decreases with an increasing $z$, $\phi_1$, and $P$. Substituting Eq. (\ref{hsnrphi}) into Eq. (\ref{esop}), we can get the the lower bound of secrecy outage probability, as shown in Eq.~(\ref{ae}).
\end{IEEEproof}

\begin{corollary}[Single-antenna Alice case] \label{singlealice}
For the case of $N_t = 1$, $P_{\text{out}}$ can be simplified to
\begin{flalign}\label{esopca1}
P_{\text{SAT}}&=1-F_{X}(\phi_2)=\frac{1}{\Gamma(N_e)}\Gamma\bigg(N_e,\frac{\phi_2}{\beta^2+|\bm{\Phi}\mathbf{h}_0|^2}\bigg),
\end{flalign}
where $\mathbf{h}_0\in \mathbb{C}^{N_s}$ is the the channel between single-antenna Alice and RIS, $\phi_2= \sigma_e^2(2^{C_{m}'-R_s}-1)/P$, $C_m'$ is defined as
\begin{flalign}\label{cmsa}
C_m'=\log_2\bigg(1+\frac{P}{\sigma^2}|\alpha \mathbf{h}_b+\mathbf{G}_r\bm{\Phi}\mathbf{h}_0|^2\bigg),
\end{flalign}
and $\mathbf{h}_b\in \mathbb{C}^{N_r\times 1}$ is the channel between single-antenna Alice and Bob. Note that $C_m'$ is achievable at Bob when using the MRC vector, i.e., $\mathbf{w}_r'=(\alpha \mathbf{h}_b+\mathbf{G}_r\bm{\Phi}\mathbf{h}_0)^{\rm{H}}|\alpha \mathbf{h}_b+\mathbf{G}_r\bm{\Phi}\mathbf{h}_0|$. $F_{X}(\phi_2)$ is defined in Eq. (\ref{cdfw1}), $|(\beta\mathbf{h}_e+\mathbf{G}_e\bm{\Phi}\mathbf{h}_0)|^2\sim X(\beta,N_e,N_s,\bm{\Phi}\mathbf{h}_0)$ represents a random variable, and $\mathbf{h}_e\sim\mathcal{CN}_{N_e,1}(\mathbf{0},\mathbf{I}_{N_e})$ is the channel from single-antenna Alice to Eve.
\end{corollary}

\begin{IEEEproof}
In the case of single-antenna Alice, beamforming vector $\mathbf{w}$ is not existed. Replacing $\mathbf{H}_e$ and $\mathbf{H}$ of Theorem 1 with $\mathbf{h}_e$ and $\mathbf{h}_0$, respectively, we can obtain Corollary \ref{singlealice}.
\end{IEEEproof}

\vspace{0.13in}

\begin{corollary}[Single-antenna Bob case] 
The secrecy outage probability of the single-antenna Bob case has the similar form with Eq. (\ref{esop}) via replacing $C_m$ by $C_m''$, where
\begin{flalign}\label{cmsb}
&C_{m}''=\log_2\bigg[1+\frac{P}{\sigma^2}|(\alpha\mathbf{h}_b^{\rm{H}}+\mathbf{h}^{\rm{H}}\bm{\Phi}\mathbf{H})\mathbf{b}|^2\bigg],
\end{flalign}
$\mathbf{h}_b^{\rm{H}}\in\mathbb{C}^{1\times N_t}$ is the channel vector between Alice and single-antenna Bob, and the channel from RIS to Bob is defined as $\mathbf{h}^{\rm{H}}\in\mathbb{C}^{1\times N_s}$.
\end{corollary} 

\begin{corollary}[Single-antenna Eve case] 
For the case of $N_e=1$, $P_{\text{out}}$ can be simplified to
\begin{flalign}\label{esopc1}
P_{\text{SAE}}(R_s)&=\exp\bigg(-\frac{\phi_1}{\beta^2+|\bm{\Phi}\mathbf{Hb}|^2}\bigg).
\end{flalign}
\end{corollary}

\begin{IEEEproof}
As $\Gamma(1,\mu) =\exp(-\mu)$ and $\Gamma(1)=1$, we can get the CDF of X as $F_{X}(\phi_1) =1-\exp[-\phi_1/(\beta^2+|\bm{\Phi}\mathbf{Hb}|^2)]$ based on Eq. (\ref{cdfw1}), then, $P_{\text{out}}$ can be deduced in the single-antenna Eve case as shown in Eq. (\ref{esopc1}).
\end{IEEEproof}

\vspace{0.2in}

\section{Alternating Optimization For Secrecy Outage Probability Minimization}\label{proposed1}

According to the expression of secrecy outage probability, i.e., Eq. (\ref{esop}), the minimization of secrecy outage probability as shown in P1 is re-formulated as follows, 
\begin{flalign}
\text{P2: } &\min_{\bm{\Phi},\mathbf{w}}\frac{1}{\Gamma(N_e)}\Gamma\bigg(N_e,\frac{\phi_1}{\beta^2+|\bm{\Phi}\mathbf{Hb}|^2}\bigg),  \\
&\text{s.t. } \mathbf{b}=\mathbf{w}/\sqrt{P},  \; \text{Eqs. } (\ref{P1C1}) \text{ and } (\ref{P1C2}). \label{P2C3}
\end{flalign}
According to Corollary 1, it is concluded that $P_{\text{out}}$ decreases with increasing $z=\frac{\phi_1}{\beta^2+|\bm{\Phi}\mathbf{Hb}|^2}$ and $P$. Hence, P2 can be transformed equivalently into P3 via using total transmission power, and P3 is expressed as
\begin{flalign}
\text{P3: } &\max_{\bm{\Phi},\mathbf{b}}\frac{\phi_1}{\beta^2+|\bm{\Phi}\mathbf{Hb}|^2},  \label{P3O} \\
&\text{s.t. } \mathbf{b}^{\rm{H}}\mathbf{b}= 1, \quad P=\rho,  \quad \text{Eq. } (\ref{P1C1}). \label{P3C1}
\end{flalign}
It is obvious that the objective function of P3 is non-convex with variables $\bm{\Phi}$ and $\mathbf{b}$. We transform P3 into two subproblems as P4 and P5, i.e., the optimization problems of beamforming vector and phase shifter matrix as follows. 
\begin{flalign}
&\text{P4: } \max\limits_{\mathbf{b}}\frac{\phi_1}{\beta^2+|\bm{\Phi}\mathbf{Hb}|^2},  \notag \\
& \quad \;\ \text{s.t. } \mathbf{b}^{\rm{H}}\mathbf{b}=1, \quad P=\rho.  \\
& \text{P5: } \max\limits_{\bm{\Phi}}\frac{\phi_1}{\beta^2+|\bm{\Phi}\mathbf{Hb}|^2},  \notag \\
& \quad \;\ \text{s.t. } \text{Eq. } (\ref{P1C1}). 
\end{flalign}
P4 is optimally solved by a closed-form solution in Section V. A, and this paper provides two optimization algorithms to solve P5, namely, semidefinite relaxation (SDR)-based method and manifold-based method in Section V. B and V. C, respectively. Then, we use an AO algorithm to find the global results for $\bm{\Phi}$ and $\mathbf{b}$ of the original problem P3.  

\subsection{Closed-Form Optimal Beamforming Vector}
For any given phase shifter matrix $\bm{\Phi}$ and $P=\rho$, the objective function of P4 is expressed as follows.
\begin{flalign}\label{p4}
\frac{\phi_1}{\beta^2+|\bm{\Phi}\mathbf{Hb}|^2}=c\bigg(\frac{\mathbf{b}^{\rm{H}}\mathbf{A}_1\mathbf{b}+t}{\mathbf{b}^{\rm{H}}\mathbf{A}_2\mathbf{b}+\beta^2}\bigg), 
\end{flalign}
where $c=\sigma_e^2/(\sigma^2 2^{R_s})$, $t=\sigma^2(1-2^{R_s})/\rho$, $\mathbf{A}_1=(\alpha\mathbf{H}_b+\mathbf{G}_r\bm{\Phi}\mathbf{H})^{\rm{H}}(\alpha\mathbf{H}_b+\mathbf{G}_r\bm{\Phi}\mathbf{H})$, and $\mathbf{A}_2=\mathbf{H}^{\rm{H}}\mathbf{H}$. According to Eq. (\ref{p4}), P4 with the given $\bm{\Phi}$ and $P=\rho$ can be transformed as
\begin{flalign}
\text{P6: } &\max_{\mathbf{b}}c\bigg(\frac{\mathbf{b}^{\rm{H}}\mathbf{A}_1\mathbf{b}+t}{\mathbf{b}^{\rm{H}}\mathbf{A}_2\mathbf{b}+\beta^2}\bigg),\notag \\
& \text{s.t. } \mathbf{b}^{\rm{H}}\mathbf{b}=1.
\end{flalign}
We use the generalized Rayleigh quotient to solve P6 \cite{absil2009optimization}, where the optimal beamforming vector $\mathbf{b}$ is the generalized eigenvector of the matrix pencil $(\mathbf{A}_1+t\mathbf{I}_{N_t},\mathbf{A}_2+\beta^2\mathbf{I}_{N_t})$, i.e., $(\mathbf{A}_2+\beta^2\mathbf{I}_{N_t})^{-1}(\mathbf{A}_1+t\mathbf{I}_{N_t})$. In detail, the optimal $\mathbf{b}$ can be deduced as follows,
\begin{flalign}\label{opbm}
\mathbf{b}^*=\text{eigvec}_{\lambda_{\max}}[(\mathbf{A}_2+\beta^2\mathbf{I}_{N_t})^{-1}(\mathbf{A}_1+t\mathbf{I}_{N_t})], 
\end{flalign}
where $\text{eigvec}_{\lambda_{\max}}(\mathbf{X})$ means the corresponding eigenvector of the largest eigenvalue of matrix $\mathbf{X}$, and $\lambda_{\max}$ is the largest eigenvalue of $\mathbf{X}$ \cite[Pro 2.1.1]{absil2009optimization}. At last, we have the optimal beamforming vector, i.e., $\mathbf{w}^*=\sqrt{\rho}\mathbf{b}^*$. It can be said that the subproblem P4 is solved optimally.   

\subsection{Phase Shifter Matrix Optimization via SDR}
Here, we use the SDR technique to solve P5. For any given bemforming vector $\mathbf{b}$ and $P=\rho$, the objective function of P5 is given as follows,
\begin{flalign}\label{p1}
 \frac{\phi_1}{\beta^2+|\bm{\Phi}\mathbf{Hb}|^2}=c\bigg[\frac{|(\alpha\mathbf{H}_b+\mathbf{G}_r\bm{\Phi}\mathbf{H})\mathbf{b}|^2+t}{\beta^2+|\bm{\Phi}\mathbf{Hb}|^2}\bigg],
\end{flalign}
where $c=\sigma_e^2/(\sigma^2 2^{R_s})$ and $t=\sigma^2(1-2^{R_s})/\rho$. Ignoring the constant $c$, the bottom part of  Eq. (\ref{p1}) is a fixed value with given $\beta$, $\mathbf{b}$, and $\mathbf{H}$, which is expressed as 
\begin{flalign}\label{p3}
\beta^2+|\bm{\Phi}\mathbf{Hb}|^2&=\beta^2+\mathbf{b}^{\rm{H}}\mathbf{H}^{\rm{H}}\bm{\Phi}^{\rm{H}}\bm{\Phi}\mathbf{Hb}\notag \\
&=\beta^2+\mathbf{b}^{\rm{H}}\mathbf{H}^{\rm{H}}\mathbf{Hb}.
\end{flalign}
To address the top part of Eq. (\ref{p1}), $\mathbf{G}_r\bm{\Phi}\mathbf{H}\mathbf{b}$ is re-formulated as \cite{Wuirs2019}
\begin{flalign}\label{transm}
\mathbf{G}_r\bm{\Phi}\mathbf{H}\mathbf{b}=\mathbf{G}_r\text{diag}(\mathbf{H}\mathbf{b})\mathbf{q}=\bm{\Sigma}\mathbf{q},
\end{flalign}
where $\mathbf{q}=\text{vec}(\bm{\Phi})=[\exp(j\theta_1),..., \exp(j\theta_{N_s})]^{\rm{T}}$. By using Eq. (\ref{transm}) and ignoring the constant $c$, the top part of Eq. (\ref{p1}) can be transformed as follows,
\begin{flalign}\label{p2}
& |(\alpha\mathbf{H}_b+\mathbf{G}_r\bm{\Phi}\mathbf{H})\mathbf{b}|^2+t \notag \\
&=\mathbf{q}^{\rm{H}}\bm{\Sigma}^{\rm{H}}\bm{\Sigma}\mathbf{q}+\alpha^2|\mathbf{H}_b\mathbf{b}|^2+\alpha\mathbf{b}^{\rm{H}}\mathbf{H}_b^{\rm{H}}\bm{\Sigma}\mathbf{q}+\alpha \mathbf{q}^{\rm{H}}\bm{\Sigma}^{\rm{H}}\mathbf{H}_b\mathbf{b}+t \notag \\
&=\hat{\mathbf{q}}^{\rm{H}}\mathbf{W}\hat{\mathbf{q}}+\alpha^2|\mathbf{H}_b\mathbf{b}|^2+t,
\end{flalign}
where 
\begin{flalign}
\mathbf{W}=
\left[
\begin{matrix} \label{reell}
\bm{\Sigma}^{\rm{H}}\bm{\Sigma}& \alpha\bm{\Sigma}^{\rm{H}}\mathbf{H}_b\mathbf{b}\\
\alpha\mathbf{b}^{\rm{H}}\mathbf{H}_b^{\rm{H}}\bm{\Sigma} & 0
\end{matrix}
\right], \; \hat{\mathbf{q}}^{\rm{H}}=\left[
\begin{matrix}
\mathbf{q}^{\rm{H}}, & l
\end{matrix}
\right],
\end{flalign}
and $l$ is an auxiliary variable. According to Eqs. (\ref{p3}) and (\ref{p2}), P5 with given $\mathbf{b}$ and $P=\rho$ can be transformed as
\begin{flalign}
\text{P7: } &\max_{\hat{\mathbf{q}}} \hat{\mathbf{q}}^{\rm{H}}\mathbf{W}\hat{\mathbf{q}}+\alpha^2|\mathbf{H}_b\mathbf{b}|^2+t, \label{P7O} \\
&\text{s.t. } |\hat{q}_i|^2=1, i=1,...,N_s+1, \label{P7C1}
\end{flalign}
where $\hat{q}_i$ is the $i$-th element in $\hat{\mathbf{q}}$. P7 is a Boolean quadratic program belonging to the NP-hard problem \cite{Luo2010SemidefiniteRelaxationQuadratic}. To address P7, we use the method in \cite{Qiao2020} that introduces $N_s+1$ auxiliary matrices $\mathbf{E}_n,n=1,...,N_s+1$ as follows,
\begin{equation}
[\mathbf{E}_n]_{i,j}=
\begin{cases}
1,&\mbox{$i=j=n$},\\
0, &\mbox{otherwise},
\end{cases}
\end{equation}
where $[\mathbf{E}_n]_{i,j}$ is the $(i,j)$-th element of $\mathbf{E}_n$. With $\mathbf{E}_n,\forall n$, the constraint (\ref{P7C1}) can be transformed as $\hat{\mathbf{q}}^{\rm{H}}\mathbf{E}_n\hat{\mathbf{q}}=1, \forall n$. Then, P7 can be transformed equivalently as 
\begin{flalign}
\text{P8: } &\max_{\mathbf{Q}}\text{tr}(\mathbf{W}\mathbf{Q})+\alpha^2|\mathbf{H}_b\mathbf{b}|^2+t,  \label{P8O} \\
&\text{s.t. }   \text{tr}(\mathbf{E}_n\mathbf{Q})=1, \forall n,  \label{P8C1}\\
& \quad \;\; \text{rank}(\mathbf{Q})=1, \mathbf{Q}\succeq \mathbf{0}, \label{P8C2}
\end{flalign}
where $\mathbf{Q}=\hat{\mathbf{q}}\hat{\mathbf{q}}^{\rm{H}}$. The objective function and constraints in P8 are convex except that $\text{rank}(\mathbf{Q})=1$ is a non-convex constraint. Thus, we use the SDR technique to drop this rank-one constraint to get a convex problem, i.e., P8 without the rank-one constraint can be optimally solved by convex optimization tools, such as interior-point methods or the MATLAB CVX tool, outputting optimal $\mathbf{Q}$, i.e., $\mathbf{Q}^*$. However, there is no guarantee that $\mathbf{Q}^*$ is the desired rank-one solution, so the Gaussian randomization method or maximum eigenvalue method is used to obtain the near-optimal $\hat{\mathbf{q}}$, as well as the near-optimal $\bm{\Phi}$ \cite{Luo2010SemidefiniteRelaxationQuadratic}. To reduce the computational cost, we chose the maximum eigenvalue method because P8 usually yields rank-one solutions (higher than 99\% of the tested cases in the investigation \cite{Wang2014}). In detail, we apply the eigenvalue decomposition on $\mathbf{Q}^*$ to get $\mathbf{Q}^*=\sum_{i=1}^{N_s+1}\lambda_i\mathbf{\bar{q}}_i\mathbf{\bar{q}}_i^{\rm{H}}$. Then, the near-optimal $\hat{\mathbf{q}}=\mathbf{\bar{q}}_1$, i.e., the eigenvector of the largest eigenvalue $\lambda_1$ of $\mathbf{Q}^*$. At last, we get the near-optimal $\bm{\Phi}=\text{diag}(\mathbf{q}_1)$ where $\mathbf{q}_1$ is $\mathbf{\bar{q}}_1$ after removing the last element.

If using the SDR-based approach for the $\bm{\Phi}$ generation, the number of variables $\mathbf{Q}$ of P8 is $N_s+1$, and P8 has one linear matrix inequality (LMI) constraint with the size of $N_s+1$. In this case, the iterations of the interior-point method are $\ln(1/\epsilon)[2(N_s+1)^{4.5}+(N_s+1)^{3.5}]$, where $\epsilon$ is the accuracy requirement of the interior-point method \cite{Wang2014,Luo2010SemidefiniteRelaxationQuadratic}. In addition, the maximum eigenvalue method requires $(N_s+1)^3+N_s$ iterations. In this case, the number of iterations in the SDR-based approach is $\ln(1/\epsilon)[2(N_s+1)^{4.5}+(N_s+1)^{3.5}]+(N_s+1)^3+N_s$.

\subsection{Phase Shifter Matrix Optimization via Manifolds}
Here, we provide another optimization algorithm for the phase shifter matrix based on manifold optimization \cite{manopt}, which can handle the unit modulus constraint with lower computational complexity. At first, according to Eq. (\ref{p2}), the minus of the objective function in P5 can be expressed as a function with respect to $\mathbf{q}$, i.e.,
\begin{flalign}\label{mo1}
 f(\mathbf{q})&=\frac{-\phi_1}{\beta^2+|\bm{\Phi}\mathbf{Hb}|^2}  \notag \\
 &=-k(\mathbf{q}^{\rm{H}}\bm{\Sigma}^{\rm{H}}\bm{\Sigma}\mathbf{q}+\alpha^2|\mathbf{H}_b\mathbf{b}|^2 +\alpha\mathbf{b}^{\rm{H}}\mathbf{H}_b^{\rm{H}}\bm{\Sigma}\mathbf{q}\notag \\
 &+\alpha \mathbf{q}^{\rm{H}}\bm{\Sigma}^{\rm{H}}\mathbf{H}_b\mathbf{b}+t),
\end{flalign}
where $k=c/(\beta^2+\mathbf{b}^{\rm{H}}\mathbf{H}^{\rm{H}}\mathbf{Hb})$ due to Eqs. (\ref{p1}) and (\ref{p3}). Then, the constraint (\ref{P1C1}) is defined a complex circle manifold as
\begin{flalign}\label{oman}
\mathcal{O}=\big\{\mathbf{q}\in\mathbb{C}^{N_s}\big||q_i|^2=1,i=1,...,N_s\big\}.
\end{flalign}
It is obvious that Eq. (\ref{oman}) is equivalent to constraint (\ref{P1C1}), and the optimal point, i.e., $\mathbf{q}^*$ is on the complex circle manifold $\mathcal{O}$. In this case, P5 with the given $\bm{\Phi}$ and $P=\rho$ can be equivalently transformed as
\begin{flalign}
\text{P9: } &\min_{\mathbf{q}\in\mathcal{O}}f(\mathbf{q}). \label{P9O} 
\end{flalign}
P9 can be solved by manifold optimization. Similar to traditional optimization methods, manifold optimization is also based on the gradient descent criterion, called the Riemannian gradient descent. However, the gradient of a point of the manifold is decided jointly by the Euclidean gradient and the tangent space on manifold $\mathcal{O}$ at this point. For instance, the tangent space for $\mathcal{O}$ at the point of the $j$th iteration, i.e., $\mathbf{q}_j\in \mathcal{O}$, is expressed as
\begin{flalign}\label{ts}
T_{\mathbf{q}_j}\mathcal{O}=\big\{\mathbf{v}\in\mathbb{C}^{N_s}\big| \mathfrak{R}(\mathbf{v}\circ\mathbf{q}_j^{\rm{T}})=\mathbf{0}\big\},
\end{flalign}
where $\mathbf{v}$ is the tangent vector at $\mathbf{q}_j$, $\circ$ is the Hadamard product, and $\mathfrak{R}(\cdot)$ means the real part of a complex variable. Among all tangent vectors on $T_{\mathbf{q}_j}\mathcal{O}$, the one that yields the fastest increase of the objective function is defined as the Riemannian gradient \cite{Yuiccc2019}, i.e., $\text{grad}_{\mathbf{q}_j}f(\mathbf{q})$, where $f(\mathbf{q})$ is the objective function with respect to $\mathbf{q}$ as defined in Eq. (\ref{mo1}). $\text{grad}_{\mathbf{q}_j}f(\mathbf{q})$ is the projection from the Euclidean gradient, i.e.,
\begin{flalign}
\triangledown_{\mathbf{q}_j}f(\mathbf{q})=-2k\bm{\Sigma}^{\rm{H}}\bm{\Sigma}\mathbf{q}_j-2k\alpha \bm{\Sigma}^{\rm{H}}\mathbf{H}_b\mathbf{b},
\end{flalign}
to the tangent space $\mathcal{O}$ as
\begin{flalign}\label{grad}
& \text{grad}_{\mathbf{q}_j}f(\mathbf{q})=\triangledown_{\mathbf{q}_j}f(\mathbf{q})-\mathfrak{R}(\triangledown_{\mathbf{q}_j}f(\mathbf{q})\circ \mathbf{q}_j^{\dagger})\circ\mathbf{q}_j,
\end{flalign}
where $(\cdot)^{\dagger}$ is the conjugate operation. The next point $\mathbf{q}_{j+1}$ should be with the direction of $\eta_j\mathbf{p}_j$ where $\eta_j$ is the step size and $\mathbf{p}_i$ is the search direction vector. The initial direction $\mathbf{p}_0$ is $\text{grad}_{\mathbf{q}_0}f(\mathbf{q})$ assuming that $\mathbf{q}_0$ is the beginning point. However, the movement can not guarantee that point $\mathbf{q}_{j+1}$ is on manifold $\mathcal{O}$. Hence, we introduce the retraction function to map a vector on $T_{\mathbf{q}_j}\mathcal{O}$ onto manifold $\mathcal{O}$, which is given as
\begin{flalign}\label{next}
\mathbf{q}_{j+1}=\text{R}_{\mathbf{q}_j}(\eta_j\mathbf{p}_j),
\end{flalign}
where a typical retraction is the normalization function, i.e., $\text{R}_{\mathbf{x}}(\mathbf{y})=\frac{y_i}{|y_i|},\forall i$. Based on the Riemannian gradient and retraction function, we use the conjugate-gradient descent method to find the optimal phase shifter matrix, which is shown in Algorithm \ref{psmo}. In conjugate-gradient descent algorithm, the update rule for the search direction on manifolds is given by
\begin{flalign}\label{sd}
\mathbf{p}_{j+1}=-\text{grad}_{\mathbf{q}_{j+1}}f(\mathbf{q})+\varphi_{j}\mathcal{T}_{\mathbf{q}_{j}\to\mathbf{q}_{j+1}}(\mathbf{p}_{j}),
\end{flalign}
where $\mathcal{T}_{\mathbf{q}_{j}\to\mathbf{q}_{j+1}}(\mathbf{p}_{j})$ is the mapping function of the tangent vector $\mathbf{p}_{j}$ from the tangent space $T_{\mathbf{q}_{j}}\mathcal{O}$ to the tangent space $T_{\mathbf{q}_{j+1}}\mathcal{O}$. The mapping function is given as
\begin{flalign}
\mathcal{T}_{\mathbf{q}_{j}\to\mathbf{q}_{j+1}}(\mathbf{p}_{j})=\mathbf{p}_{j}-\mathfrak{R}(\mathbf{p}_{j}\circ\mathbf{q}_{j+1}^{\dagger})\circ\mathbf{q}_{j+1}.
\end{flalign}

Since the second derivative of $f(\mathbf{q})$ is $\triangledown^2_{\mathbf{q}}f(\mathbf{q})=-k\bm{\Sigma}^{\rm{H}}\bm{\Sigma}$ and $\bm{\Sigma}^{\rm{H}}\bm{\Sigma}$ is positive semidefinite, it is concluded that $f(\mathbf{q})$ is concave, and Algorithm \ref{psmo} based on conjugate-gradient descent can not be guaranteed to converge to the optimal point  \cite{manopt}. The output $\bm{\Phi}^*$ in Algorithm \ref{psmo} is the local optimal result.

The computational complexity analysis of Algorithm \ref{psmo} is discussed here. The conjugate-gradient descent algorithm needs $N_s^2$ iterations for convergence \cite{manopt}, and each iteration needs the calculations of Eqs. (\ref{grad}), (\ref{next}), (\ref{sd}), and a step size update, i.e., the steps 4-7. The steps 4, 5, 6, and 7 require $4N_s^2+2N_s$, $2N_s$, $N_s$, and $4N_s^2$ inner-iterations, respectively. In the step 9, the generation of $\bm{\Phi}^*$ needs $N_s$ iterations. In total, the number of iterations of Algorithm \ref{psmo} is $8N_s^4+5N_s^3+N_s$.

\begin{algorithm}
\small
\KwData{$\alpha, \beta, \rho, R_s, \sigma^2, \sigma_e^2, \mathbf{b}, \mathbf{H}_b, \mathbf{G}_r, \mathbf{H}, N_s$}
\KwResult{$\bm{\Phi}^*$}
Initialize $\eta_0$ and $\varphi_0$\;
Select beginning point $\mathbf{q}_0$, calculate $\text{grad}_{\mathbf{q}_0}f(\mathbf{q})$\;
\While{$|\rm{grad}_{\mathbf{q}_j}f(\mathbf{q})|\leq \zeta $}{
Calculate Riemannian gradient $\text{grad}_{\mathbf{q}_j}f(\mathbf{q})$ via Eq. (\ref{grad})\;
Compute conjugate search direction $\mathbf{p}_{j}$ via Eq. (\ref{sd})\;
Find next point $\mathbf{q}_{j+1}$ via Eq. (\ref{next}) \;
Determine step size $\eta_j$ and $\varphi_j$ proposed in \cite{absil2009optimization} \;
}
Get $\mathbf{q}^*=\mathbf{q}_j$\;
$\bm{\Phi}^*=\text{diag}(\mathbf{q}^*)$\;
\textbf{Procedure End}
\caption{Conjugate-gradient Descent Algorithm for Phase Shifter Matrix based on Manifold.}\label{psmo}
\end{algorithm}

\subsection{Alternating Optimization}
The SDR-based or manifold-based optimization method is used to find the optimal phase shifter matrix with given beamforming vectors, and Eq. (\ref{opbm}) provides the optimal beamforming vector for given phase shifter matrices. Referring to \cite{Huang2019}, we design an  AO algorithm for the global results of the phase shifter matrix and beamforming vector, as shown in Algorithm \ref{ao}. The AO algorithm is an iterative procedure for the global optimization over $\bm{\Phi}$ and $\mathbf{b}$ by alternating restricted optimization over individual $\bm{\Phi}$ and $\mathbf{b}$. It provides global results that can converge with limited iterations \cite{bezdek2003convergence}. The following theorem is used to show the convergence condition.

\begin{theorem}[Convergence condition]\label{cc}
For given convex sub-problems, i.e., P6 and P8 removing the rank-one constraint, there exists a global convergence point $\bm{\Omega}_{\rm{iter}}=(\mathbf{b}^*_{\rm{iter}}, \mathbf{Q}^*_{\rm{iter}})$ that is the output of the $\rm{iter}$-th iteration, resulting in that $P_{\text{out}}(\bm{\Omega}_{\rm{iter}})$ is equal to $P_{\text{out}}(\bm{\Omega}_{\rm{iter}+1})$, i.e., the secrecy outage probability $P_{\text{out}}$ approaches to a constant, where $P_{\text{out}}(\bm{\Omega})$ is the secrecy outage probability function with parameters $\bm{\Omega}=(\mathbf{b}^*,\mathbf{Q}^*)$ that is calculated by step 5 in Algorithm \ref{ao}.

\end{theorem}

\begin{IEEEproof}
See \cite[Th. 3]{bezdek2003convergence}. In brief, if all sub-problems of the original problem are convex, there exists a global convergence point that can be found within the finite number of alternations.
\end{IEEEproof}

\begin{algorithm}
\small
\KwData{$\alpha, \beta, \rho, R_s, \sigma^2, \sigma_e^2, \mathbf{H}_b, \mathbf{G}_r, \mathbf{H}, N_t, N_s, N_e$}
\KwResult{$\bm{\Phi}_o^{*}, \mathbf{w}_o^*$}
Initialize $\rm{iter}=1$ and the iterating limit is $\textrm{iter}_{\max}$\;
Initialize beamforming vector $\mathbf{b}_0=\text{random\_unitary}(N_t,1)$\;
\While{$\rm{iter}\leq \rm{iter}_{\max}\&\&\{P_{\rm{out}}(\bm{\Omega}_{\rm{iter}+1})-P_{\rm{out}}(\bm{\Omega}_{\rm{iter}})\leq \xi\}$}{
Solve P8 via interior point method to obtain $\mathbf{Q}_{\rm{iter}}$\;
Calculate $P_{\text{out}}(\bm{\Omega}_{\rm{iter}})$ via Eqs. (\ref{esop}), (\ref{p1}), and (\ref{P8O})\;
Solve P4 via Eq. (\ref{opbm}) to obtain $\mathbf{b}_{\rm{iter}+1}$\;
}

$\mathbf{Q}_o^{*}=\mathbf{Q}_{\rm{iter}}, \mathbf{w}_o^*=\sqrt{\rho}\mathbf{b}_{\rm{iter}}$\;
Get $\hat{\mathbf{q}}^*_{o}$ via maximum eigenvalue method of $\mathbf{Q}_o^{*}$\;
Get $\mathbf{q}_{o}^*$ via removing the last element of $\hat{\mathbf{q}}^*_{o}$ and calculate $\bm{\Phi}^*_{o}=\text{diag}(\mathbf{q}^*_{o})$\;
\textbf{Procedure End}
\caption{AO Algorithm for Phase Shifter and Beamforming.} \label{ao}
\end{algorithm}

The global convergence point $\bm{\Omega}_{\rm{iter}}$ is the output of the alternating algorithm, where $\mathbf{Q}_o^{*}=\mathbf{Q}_{\rm{iter}}, \mathbf{w}_o^*=\sqrt{\rho}\mathbf{b}_{\rm{iter}}$ are seen as the desired results of global optimization. According to the convergence condition in Theorem \ref{cc}, we set an arbitrarily small value $\xi$, where the iteration process continues until $\{P_{\text{out}}(\bm{\Omega}_{\rm{iter}+1})-P_{\text{out}}(\bm{\Omega}_{\rm{iter}})\leq \xi\}$. To avoid the endless loop, we set $\rm{iter}_{\max}$ in Algorithm \ref{ao} as the maximum number of allowed loops. $\text{random\_unitary}(N_t,1)$ is to randomly generate an $N_t\times 1$ unitary vector in this algorithm. 

It is worth noting that the manifold-based method can not provide a provable global-optimal phase shifter matrix. Hence, the AO algorithm between P4 and P9 is not guaranteed to be convergent. Although the AO algorithm between P4 and P9 can be performed via replacing $\bm{\Omega}_{\rm{iter}}=(\mathbf{b}^*_{\rm{iter}}, \mathbf{Q}^*_{\rm{iter}})$ with $\bm{\Omega}_{\rm{iter}}=(\mathbf{b}^*_{\rm{iter}}, \bm{\Phi}^*_{\rm{iter}})$ in Algorithm \ref{ao}, the convergence performance should be examined by simulations in Section VII.

\subsection{Computational Complexity Analysis}
The computational complexity analysis of Algorithm \ref{ao} is discussed as follows. The step 4 needs $\ln(1/\epsilon)[2(N_s+1)^{4.5}+(N_s+1)^{3.5}]$ iterations. According to Eqs. (\ref{esop}), (\ref{p1}), and (\ref{P8O}), we know that the step 5 requires $N_s^2N_t+N_sN_t$ iterations. According to Eq. (\ref{opbm}), we know that the step 6 requires $3N_t^3+2N_t^2$ iterations. At last, steps 8-9 need $(N_s+1)^3+N_s$ iterations. In total, the computational complexity of Algorithm \ref{ao} in the worst condition is on the order of $O\{\text{iter}_{\max}[\ln(1/\epsilon)N_s^{4.5}+N_t^3+N_s^2N_t]\}$ where $\rm{iter}_{\max}$ is the maximum number of allowed loops. If Algorithm \ref{ao} uses the manifold-based approach for the $\bm{\Phi}_{\rm{iter}}$ generation, the computational complexity is $O[\text{iter}_{\max}(N_s^4+N_t^3+N_s^2N_t)]$. It is concluded that the AO algorithm with manifold optimization has less computational complexity than that with the SDR approach.

\section{Minimization of Secrecy Outage Probability For Single-Antenna Cases}\label{scase}

\subsection{Single-Antenna Alice}
Corollary 2 provides the expression of secrecy outage probability when Alice has one antenna. According to Eq. (\ref{esopca1}) in Corollary 2 and the monotonically decreasing of the secrecy outage probability, the minimization of secrecy outage probability, i.e., P2, can be expressed as
\begin{flalign}
\text{P10: } &\max_{\bm{\Phi}}\frac{\phi_2}{\beta^2+|\bm{\Phi}\mathbf{h}_0|^2}, \notag \\
&\text{s.t. } \text{Eq. } (\ref{P1C1}), \label{P3C1}
\end{flalign}
where $\phi_2= \sigma_e^2(2^{C_{m}'-R_s}-1)/P$. The minus of the objective function in P10 can be transformed as a function with respect to $\mathbf{q}$, i.e., 
\begin{flalign}\label{fun0}
 f_0(\mathbf{q})&=\frac{-\phi_2}{\beta^2+|\bm{\Phi}\mathbf{h}_0|^2}\notag \\
 &=-k[|\alpha\mathbf{h}_b+\mathbf{G}_r\bm{\Phi}\mathbf{h}_0|^2+t]\notag \\
&=-k(\mathbf{q}^{\rm{H}}\bm{\Sigma}_0^{\rm{H}}\bm{\Sigma}_0\mathbf{q}+\alpha^2|\mathbf{h}_b|^2\notag \\
&+\alpha\mathbf{h}_b^{\rm{H}}\bm{\Sigma}_0\mathbf{q}+\alpha \mathbf{q}^{\rm{H}}\bm{\Sigma}_0^{\rm{H}}\mathbf{h}_b+t),
\end{flalign}
where $\bm{\Sigma}_0=\mathbf{G}_r\text{diag}(\mathbf{h}_0)$, $k=c/(\beta^2+|\mathbf{h}_0|^2)$, $c=\sigma_e^2/(\sigma^2 2^{R_s})$, and $t=\sigma^2(1-2^{R_s})/\rho$. With Eq. (\ref{fun0}), P10 can be transformed as a problem of manifold optimization, i.e.,
\begin{flalign}
\text{P11: } &\min_{\mathbf{q}\in\mathcal{O}}f_0(\mathbf{q}), \label{P11O} 
\end{flalign}
where $\mathbf{q}$ is defined as a complex circle manifold as shown in Eq. (\ref{oman}). The paper also uses the conjugate-gradient descent algorithm to solve P11, whose procedure is similar to Algorithm \ref{psmo} for the multiple-antenna case, so we will not repeat it. Note that beamforming optimization is not available in the single-antenna Alice case. Thus alternating optimization is not required, and the computational complexity is reduced significantly. The computational complexity to solve P11 with the conjugate-gradient descent algorithm is on the order of $O(N_s^4)$.

\subsection{Single-Antenna Bob}

When Bob has one antenna, with given beamforming vector $\mathbf{b}$ and $P=\rho$, the minimization of secrecy outage probability, i.e., P2, can be expressed as
\begin{flalign}
\text{P12: } &\max_{\bm{\Phi}} \frac{\phi_1}{\beta^2+|\bm{\Phi}\mathbf{Hb}|^2}, \notag \\
& \text{s.t. } \text{Eq. } (\ref{P1C1}), \label{P12C1}
\end{flalign}
where $\phi_1=\sigma_e^2(2^{C_{m}''-R_s}-1)/\rho$ and $C_{m}''$ is defined in Eq. (\ref{cmsb}). The objective function of P12 can be transformed as follows,
\begin{flalign}\label{sab1}
\frac{\phi_1}{\beta^2+|\bm{\Phi}\mathbf{Hb}|^2}=k[|(\alpha\mathbf{h}_b^{\rm{H}}+\mathbf{h}^{\rm{H}}\bm{\Phi}\mathbf{H})\mathbf{b}|^2+t].
\end{flalign}
With Eq. (\ref{sab1}), P12 can be transformed as 
\begin{flalign}
\text{P13: } &\max_{\bm{\Phi}} |(\alpha\mathbf{h}_b^{\rm{H}}+\mathbf{h}^{\rm{H}}\bm{\Phi}\mathbf{H})\mathbf{b}|^2, \notag \\
& \text{s.t. } \text{Eq. } (\ref{P1C1}). \label{P12C1}
\end{flalign}
We use the method in \cite{Wuirs2019} to find the closed-form solution for the optimal phase shifter matrix $\bm{\Phi}$. At first, we have the following inequality:
\begin{flalign}
|(\alpha\mathbf{h}_b^{\rm{H}}+\mathbf{h}^{\rm{H}}\bm{\Phi}\mathbf{H})\mathbf{b}|\leq |\alpha\mathbf{h}_b^{\rm{H}}\mathbf{b}|+|\mathbf{h}^{\rm{H}}\bm{\Phi}\mathbf{H}\mathbf{b}|.
\end{flalign}
The equation holds if and only if $\text{arg}(\mathbf{h}_b^{\rm{H}}\mathbf{b})=\text{arg}(\mathbf{h}^{\rm{H}}\bm{\Phi}\mathbf{H}\mathbf{b})=\text{arg}[\mathbf{q}^{\rm{H}}\text{diag}(\mathbf{h}^{\rm{H}})\mathbf{m}]=\theta_0$, where $\mathbf{m}=\mathbf{H}\mathbf{b}$ and $\text{arg}(x)$ is the angle of the complex number $x$. Thus, with a fixed $\mathbf{b}$ and $\theta_0$, the $n$-th phase shifter at the RIS has the closed-form solution, i.e.,
\begin{flalign}\label{thetaopt}
\theta_n^*=\theta_0-\text{arg}(h^{\rm{H}}_n)-\text{arg}(m_n), n=1,...,N_s,
\end{flalign}
where $h^{\rm{H}}_n$ and $m_n$ are the $n$-th elements of $\mathbf{h}^{\rm{H}}$ and $\mathbf{m}$, respectively. According to Eq. (\ref{thetaopt}), we have $\bm{\Phi}^*=\text{diag}[\exp(j\theta_1^*),..., \exp(j\theta_{N_s}^*)]$. If $\bm{\Phi}$ is fixed, we can get the optimal beamforming vector $\mathbf{b}^*$ via Eq. (\ref{opbm}) replacing $\mathbf{A}_1$ with $\mathbf{A}_3$, where $\mathbf{A}_3=(\alpha\mathbf{h}_b^{\rm{H}}+\mathbf{h}^{\rm{H}}\bm{\Phi}\mathbf{H})^{\rm{H}}(\alpha\mathbf{h}_b^{\rm{H}}+\mathbf{h}^{\rm{H}}\bm{\Phi}\mathbf{H})$. Note that both $\mathbf{b}^*$ and $\bm{\Phi}^*$ have the closed-form solution, so the alternating optimization between beamforming and phase shifter matrix in the single-antenna Bob case has lower computational complexity compared to the multiple-antenna case. In detail, Eq. (\ref{thetaopt}) requires $N_sN_t+N_s$ iterations and the calculation of $\theta_0$ requires $N_t$ iterations, so the computational complexity of alternating optimization in the single-antenna Bob case is on the order of $O[\text{iter}_{\max}(N_t^3+N_s^2N_t)]$.

\section{Simulations} \label{simulations}
The simulation results are provided to investigate joint impacts of SNR, PLS coding rate, the number of antennas, and the number of RIS elements on the secrecy outage probability of the proposed schemes. Here, the AWGN floor parameters $\sigma^2$ and $\sigma_e^2$ are calculated by $[174+10\log2(W)+10]$ dBm, where $W$ is the carrier bandwidth and selected as 20 MHz. The path loss can be uniformly calculated by $\frac{c}{\sqrt{d_i}},i=\{1,2,3\}$, where $c$ is the path loss constant. $d_1$,  $d_2$, and $d_3$ are the distances from Alice to Bob, Alice to RIS, and RIS to Bob, respectively. $\alpha=\beta=0.8$ are statistical values and can be obtained via historical data\footnote{The performance in terms of $\alpha$ and $\beta$ can be found in \url{https://github.com/yiliangliu1990/liugit_pub/tree/master/RIS}.}

\subsection{Numerical Test for Secrecy Outage Probability}

\begin{figure}[h!]
\begin{subfigure}[t]{.45\textwidth}
\centering
\includegraphics[width=0.95\textwidth]{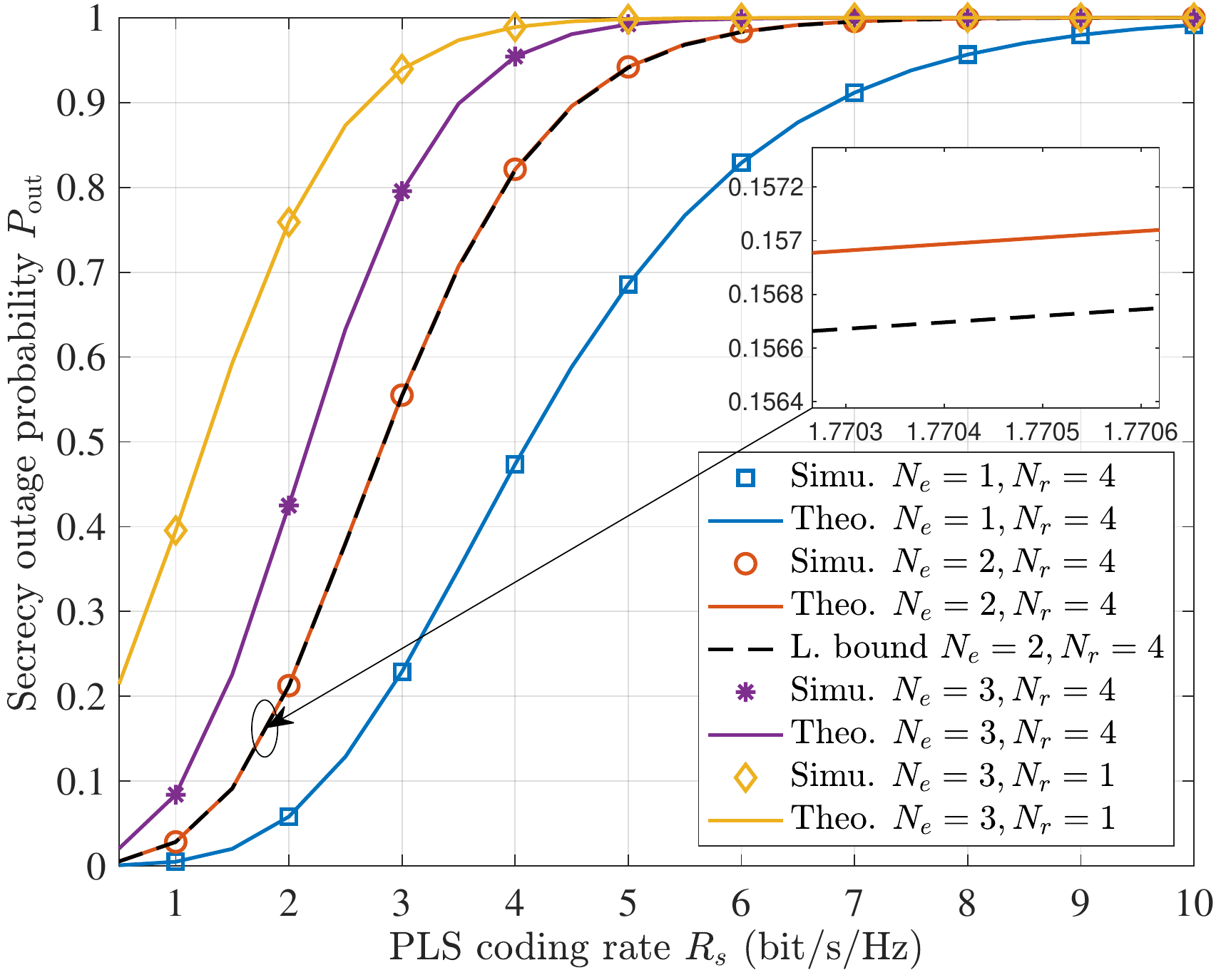}
\caption{PLS coding rate effect when SNR is 9 dB.}\label{sim1}
\end{subfigure}\hfill
\begin{subfigure}[t]{.45\textwidth}
\centering
\includegraphics[width=0.95\textwidth]{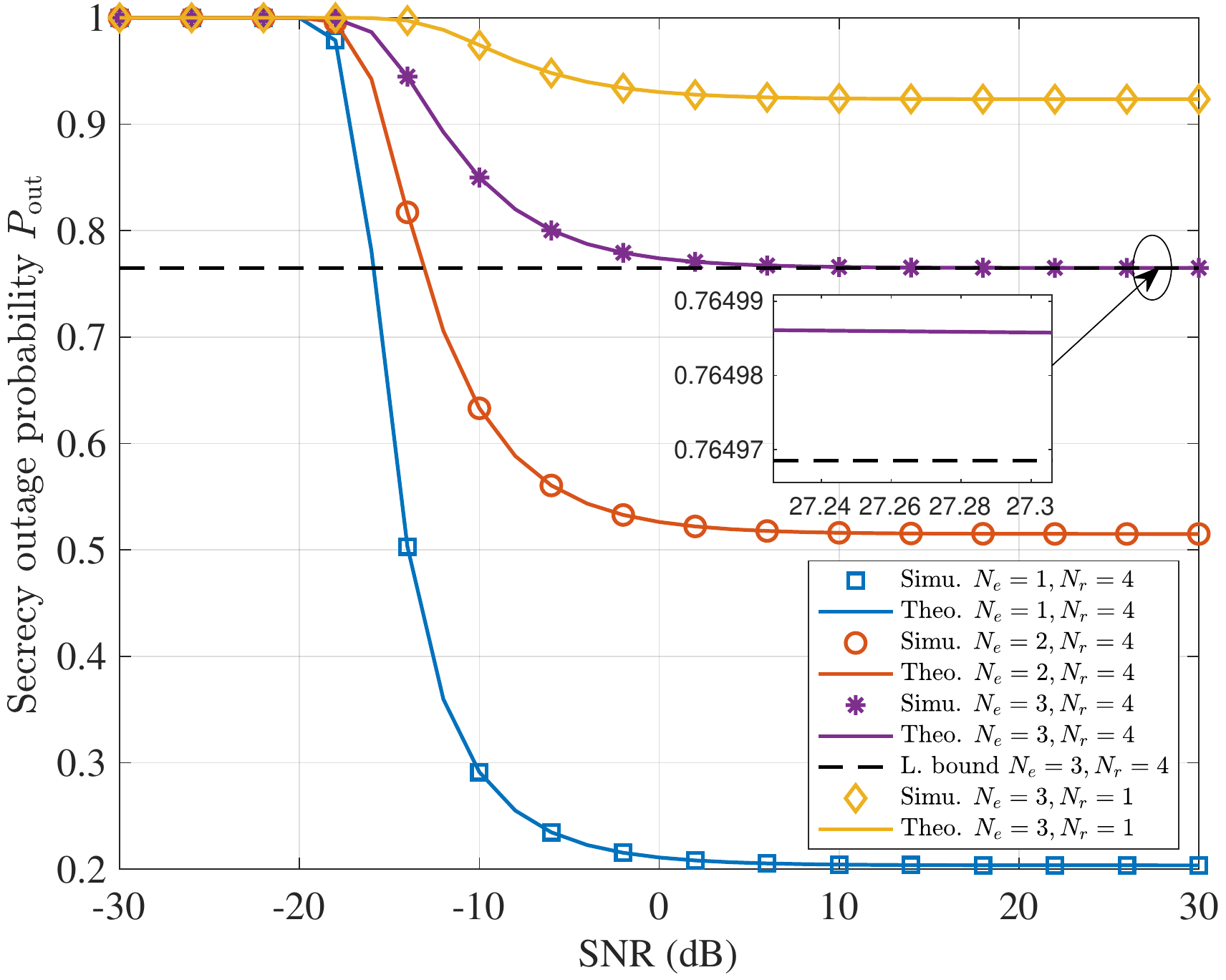}
\caption{SNR effect when PLS coding rate is 3 bit/s/Hz}\label{sim2}
\end{subfigure}
\caption{Theoretical results and Monte Carlo simulations of secrecy outage probability, where $N_s=16$, $N_t=10$, $\alpha=\beta=0.8$, $\bm{\Phi}$ is randomly generated, and $\mathbf{w}$ is the MRT vector.}
\end{figure}

Here, we examine Theorem 1 in terms of PLS coding rate and SNR in Figs. \ref{sim1} and \ref{sim2}, respectively. These figures show the good agreements between theoretical results (Theo.) and Monte Carlo simulation results (Simu.) from $10^5$ independent runs. The simulation adopts the parameters as $N_s=16$, $N_t=10$, $\bm{\Phi}$ is arbitrary, and $\mathbf{w}$ is the MRT vector of main channels, i.e., $\mathbf{w}/\sqrt{\rho}=(\alpha \mathbf{h}_b^{\rm{H}}+\mathbf{h}^{\rm{H}}\bm{\Phi}\mathbf{H})^{\rm{H}}/|\alpha \mathbf{h}_b^{\rm{H}}+\mathbf{h}^{\rm{H}}\bm{\Phi}\mathbf{H}|$ in the single-antenna Bob case, and $\mathbf{w}/\sqrt{\rho}=\text{eigvec}_{\lambda_{\max}}[(\alpha\mathbf{H}_b+\mathbf{G}_{r}\bm{\Phi}\mathbf{H})^{\rm{H}}(\alpha\mathbf{H}_b+\mathbf{G}_{r}\bm{\Phi}\mathbf{H})]$ in the multiple-antenna Bob case \cite{Kang2003}.

From Fig. \ref{sim1}, we can find that the secrecy outage probability increases with an increasing PLS coding rate. It is also shown that the increasing number of Eve's antennas enlarges the secrecy outage probability, and adding Bob's antennas can reduce secrecy outage probability, which are consistent with the conclusions \cite{Zhou2011RethinkingSecrecyOutage}. Without loss of generality, the simulations of $\{N_e=2, N_r=4\}$ are selected for the tight test between the theoretical results of secrecy outage probability and their lower bound. It is demonstrated that the lower bound from Corollary 1 is very tight for secrecy outage probability. In Fig. \ref{sim2}, we can find that the secrecy outage probability decreases fast with an increasing SNR at lower SNR regions, then decreases slowly and has a nearly constant in higher SNR regions, which is consistent with the conclusion in Corollary 1 of this paper. Throughout Figs. \ref{sim1} and \ref{sim2}, we find the secrecy outage probability with arbitrary phase shifter matrices is very large when the PLS coding rate is larger than 1 bit/s/Hz, meaning that the optimization of the phase shifter matrix is necessary.

\subsection{Single-Antenna Alice}

\begin{figure}[h]
\centering
\includegraphics[width=0.9\linewidth]{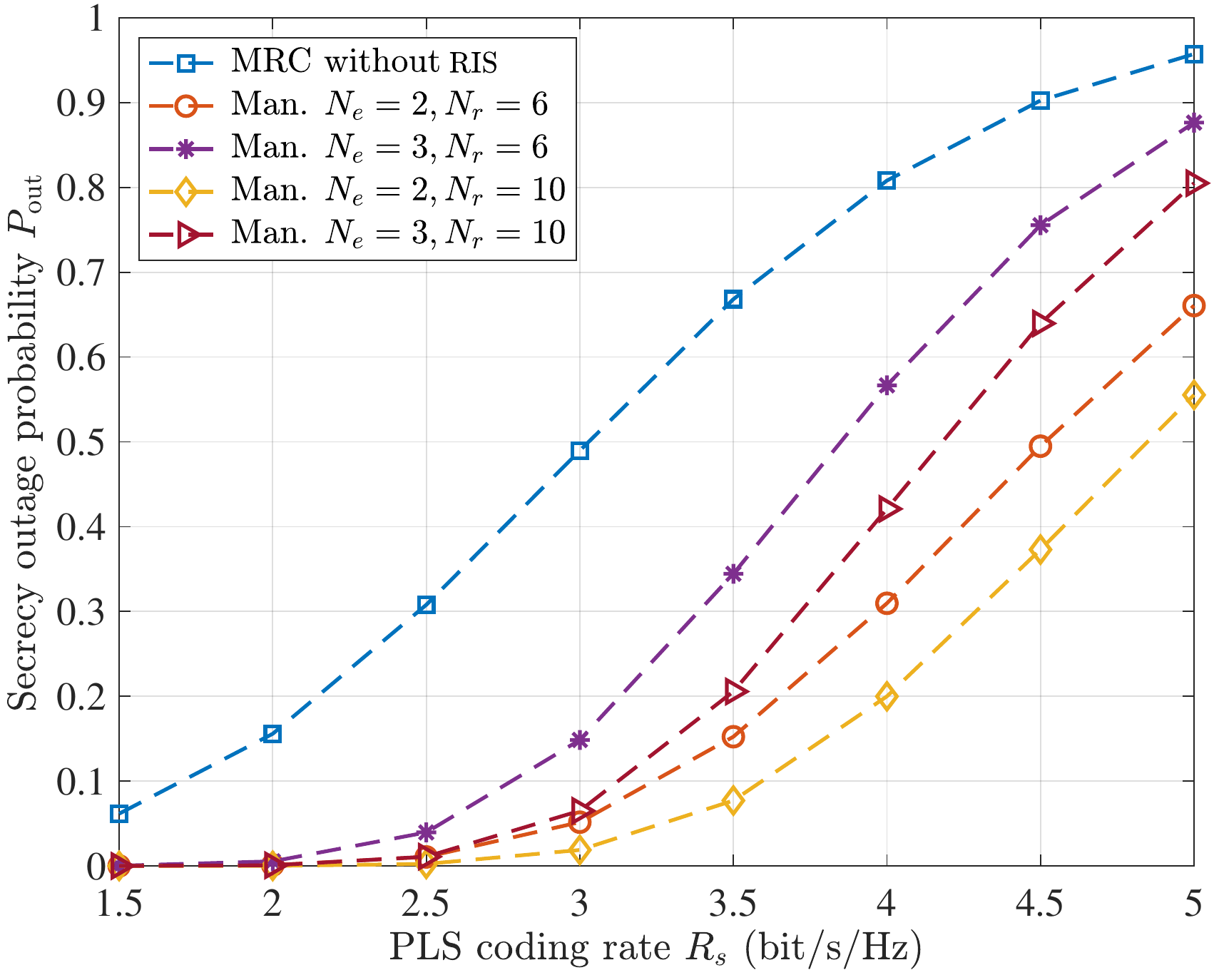}
\caption{Secrecy outage probability in RIS-assisted SIMOME, where $N_s=32$, $\alpha=\beta=0.8$, and SNR is 9 dB.}\label{simsimome}
\end{figure}

The secrecy outage probability in the case of single-antenna Alice, also called RIS-assisted single-input multiple-output multiple-antenna-eavesdropper (SIMOME) case, is examined in Fig.~\ref{simsimome}. The comparison simulation is conducted between the proposed manifold (Man.) method in Section VI. A and the MRC technology without RIS. The MRC vector without RIS is $\mathbf{h}_b^{\rm{H}}/|\mathbf{h}_b|$. We can find that an increasing PLS coding rate will enlarge the secrecy outage probability, and the proposed method has a lower secrecy outage probability compared to the MRC method without RIS. In addition, the increasing number of Eve's antennas causes a large secrecy outage probability because the wiretap channel capacity increases with the number of Eve's antennas. It is also demonstrated that adding Bob's antennas will reduce secrecy outage probability.

\subsection{Single-Antenna Bob}
In the single-antenna Bob case, three different schemes, i.e., MRT beamforming without RIS (MRT without RIS), joint MRT beamforming and phase shift optimization (MRT-PS), and the proposed AO scheme based on closed-form solutions (AO-CS) in Section VI. B, are compared:

\begin{enumerate}
\item MRT without RIS: Alice performs MRT-based beamforming, i.e., $\mathbf{w}=\sqrt{\rho}\mathbf{h}_b^{\rm{H}}/|\mathbf{h}_b|$ where $\mathbf{h}^{\rm{H}}_b$ represents the channel between Alice and Bob in the scenarios without RIS, and phase shift control is not considered in this case. The secrecy outage probability of MRT-based beamforming is calculated as $P_{\text{out,MRT}}(R_s)=\Gamma(N_e,\phi_m)/\Gamma(N_e)$, where $\phi_{m}=\sigma_e^2(2^{C_{m}-R_s}-1)/\rho$ and $C_m=\log_2(1+\rho/\sigma^2|\alpha\mathbf{h}^{\rm{H}}_b|^2)$. The computational complexity of MRT-based beamforming is $O(N_b)$.
\item MRT-PS: Alice performs joint MRT-based beamforming and SDR-based or manifold-based phase shifter optimization scheme as proposed in \cite{Wuirs2019,Hong2019,Dong2020,Wang2020}, which achieves an optimal channel capacity without the instantaneous CSIs of eavesdroppers. The secrecy outage probability of this scheme can be measured by Eq. (\ref{esop}). The computational complexity of MRT-PS is $O(N_s^4+N_sN_t+N_t)$ (using manifold optimization) or $O[\ln(1/\epsilon)(N_s^{4.5}+N_sN_t+N_t)]$ (using SDR optimization).
\item AO-CS: Alice performs the AO algorithm between beamforming and phase shifter as described in Section \ref{scase}. B. Both beamforming vectors and phase shifter matrices are closed-form in the alternating process. The secrecy outage probability of the proposed AO-CS scheme can be measured by Eq. (\ref{esop}). The computational complexity is $O\{\text{iter}_{\max}[N_t^3+N_s^2N_t]\}$ as discussed in Section \ref{scase}. E.
\end{enumerate}

\begin{figure*}[h!]
\begin{subfigure}[t]{.32\textwidth}
\centering
\includegraphics[width=1\textwidth]{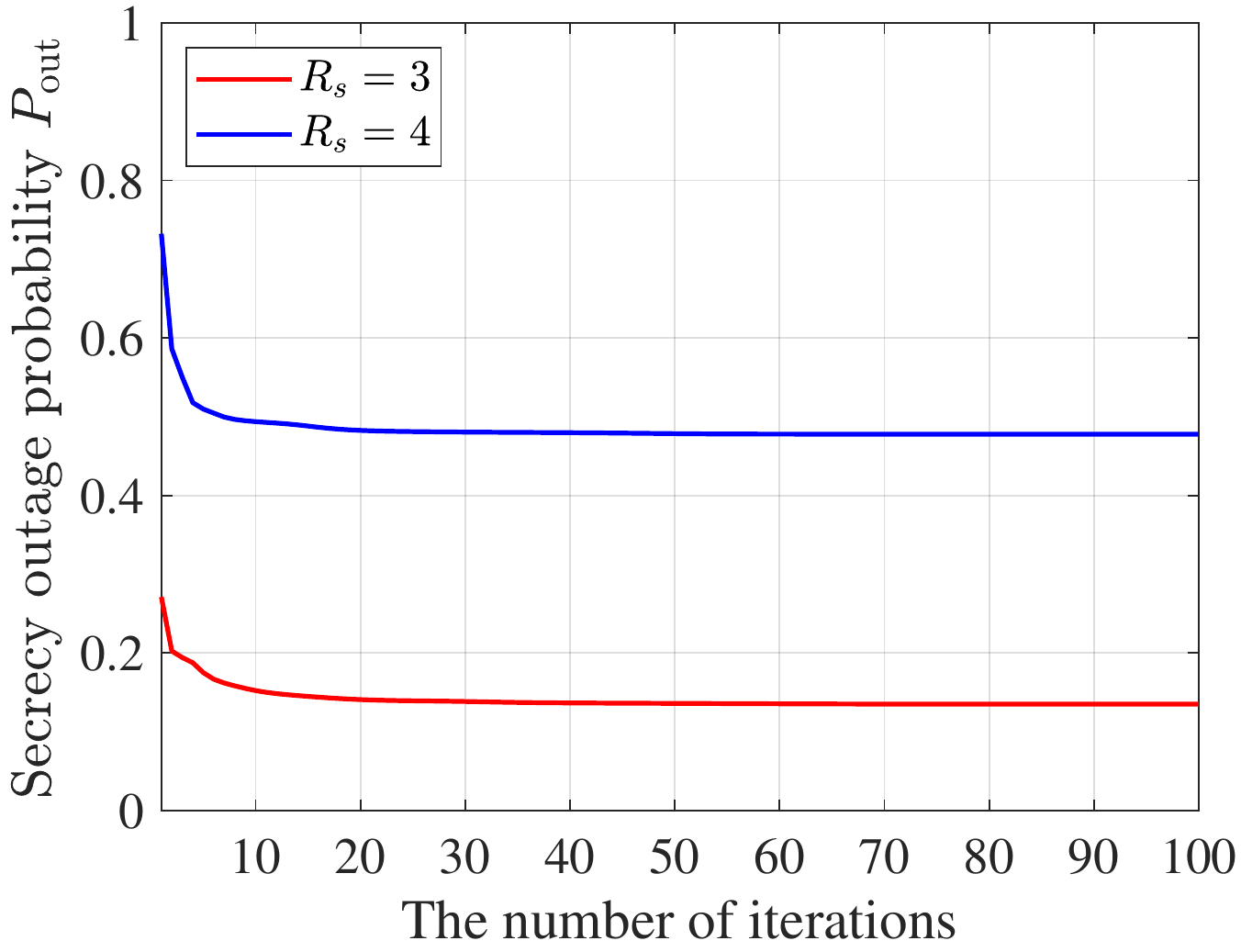}
\caption{$N_t=10$, $N_e=2$, $N_s=32$, $\alpha=\beta=0.8$, and SNR is 7 dB.}\label{convergence_sim}
\end{subfigure}\hfill
\begin{subfigure}[t]{.32\textwidth}
\centering
\includegraphics[width=1\textwidth]{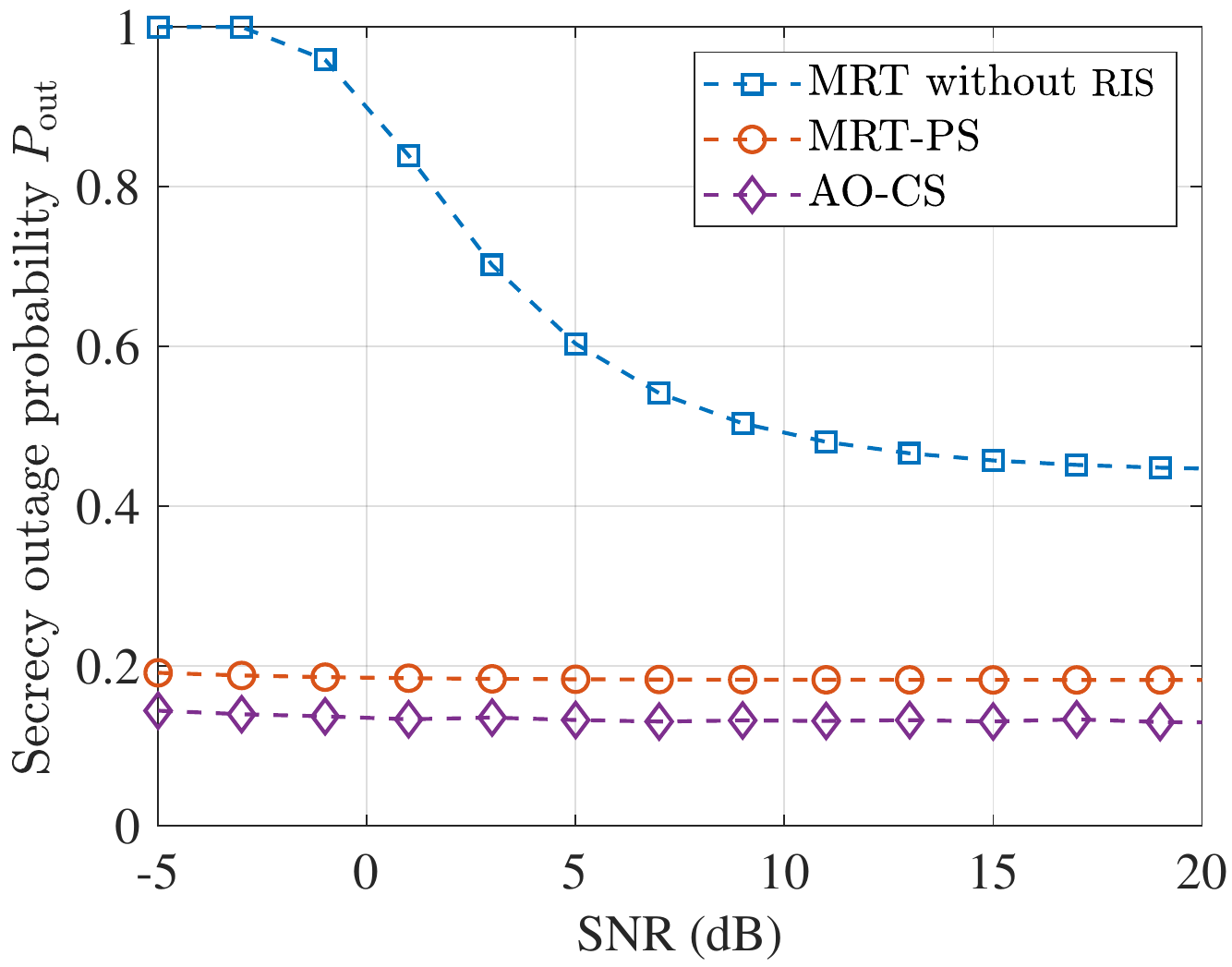}
\caption{$N_t=10$, $N_e=2$, $N_s=32$, $\alpha=\beta=0.8$, and $R_s$ is 3 bit/s/Hz.}\label{simSNR}
\end{subfigure}
\begin{subfigure}[t]{.32\textwidth}
\centering
\includegraphics[width=1\textwidth]{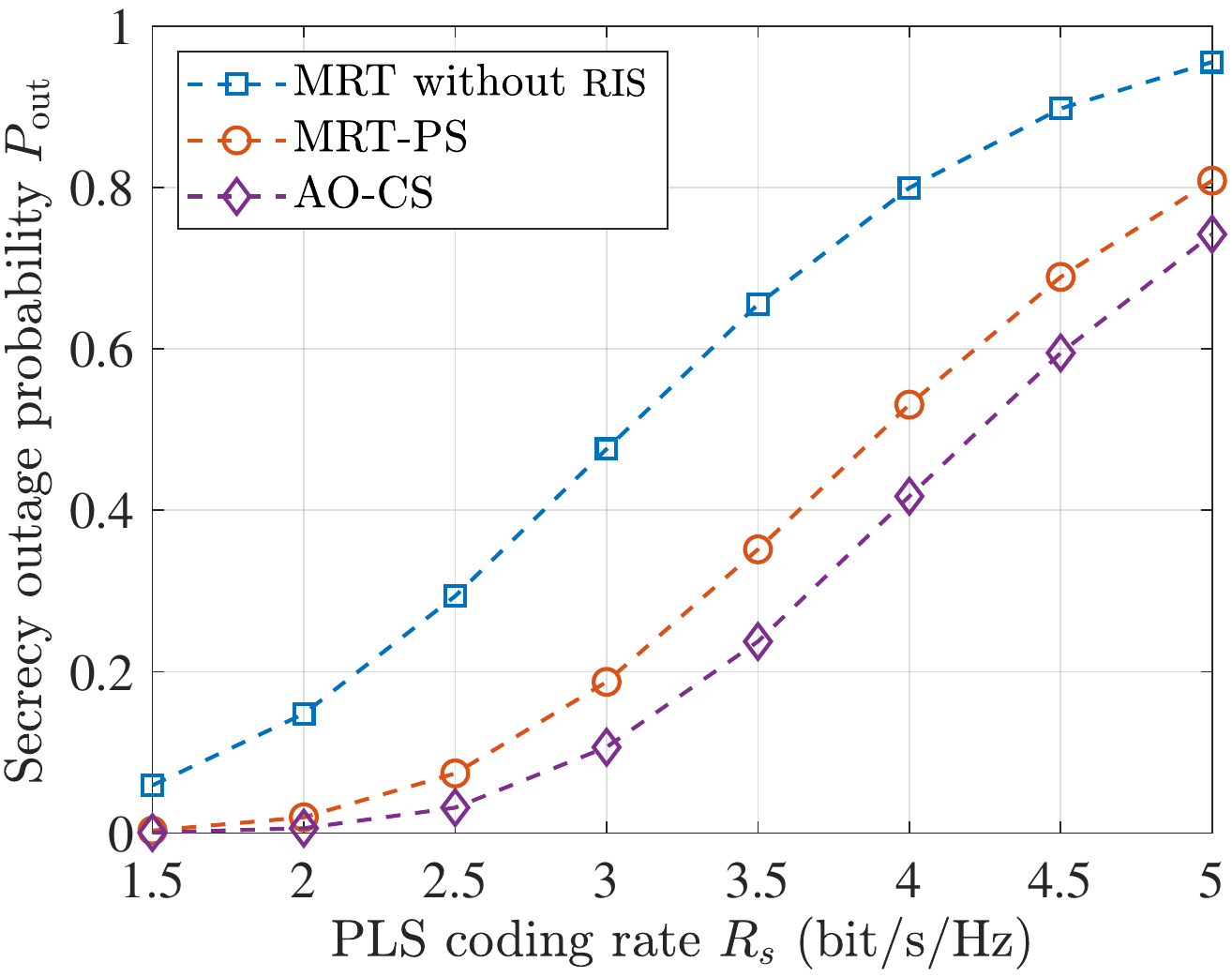}
\caption{$N_t=10$, $N_e=2$, $N_s=32$, $\alpha=\beta=0.8$, and SNR is 9 dB.}\label{simRs}
\end{subfigure}
\begin{subfigure}[t]{.32\textwidth}
\centering
\includegraphics[width=1\textwidth]{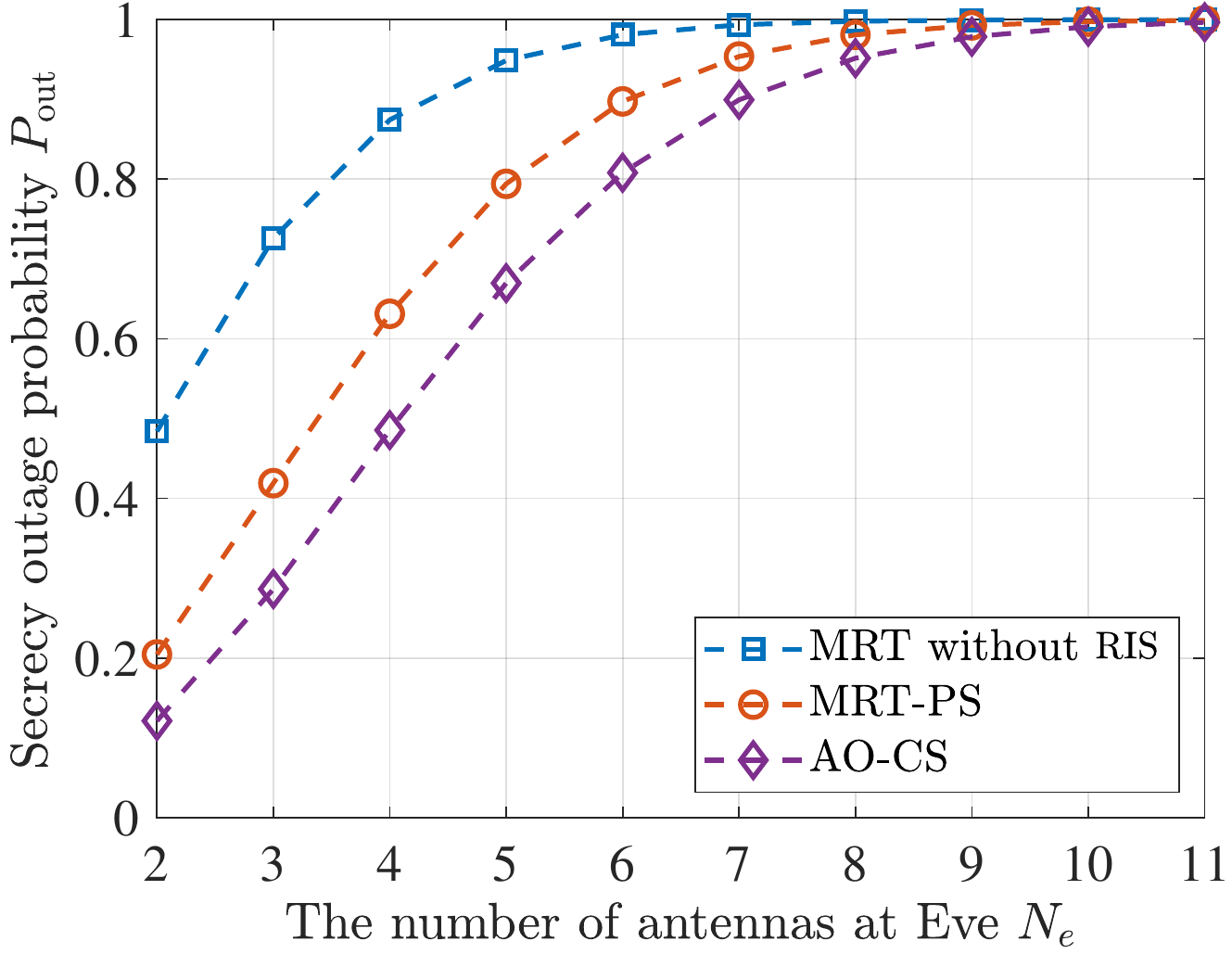}
\caption{$N_t=10$, $N_s=32$, $\alpha=\beta=0.8$, SNR is 9 dB, and $R_s=3$ bit/s/Hz.}\label{simNe}
\end{subfigure}\hfill
\begin{subfigure}[t]{.32\textwidth}
\centering
\includegraphics[width=1\textwidth]{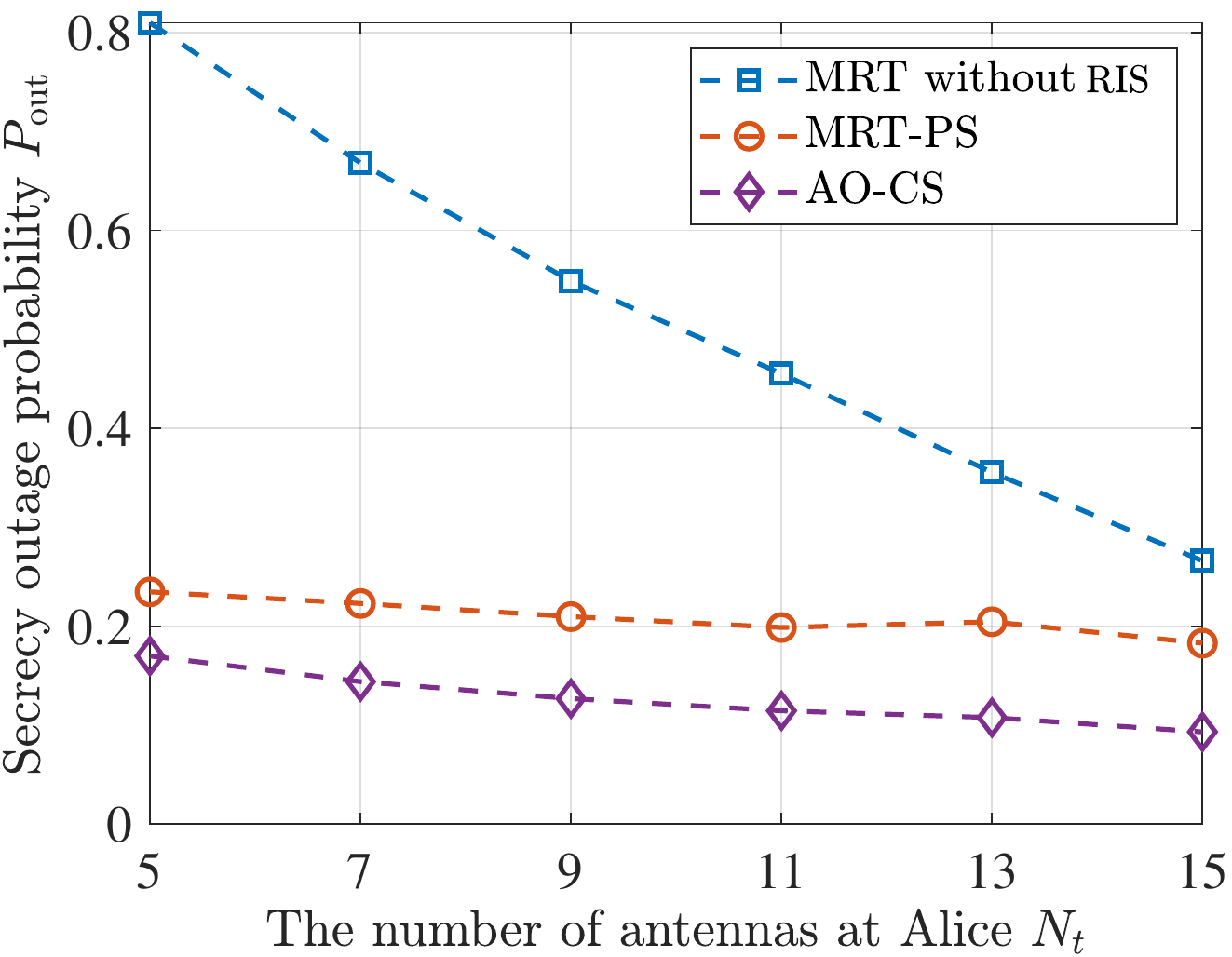}
\caption{$N_e=2$, $N_s=32$, $\alpha=\beta=0.8$, SNR is 9 dB, and $R_s=3$ bit/s/Hz.}\label{simal}
\end{subfigure}
\begin{subfigure}[t]{.32\textwidth}
\centering
\includegraphics[width=1\textwidth]{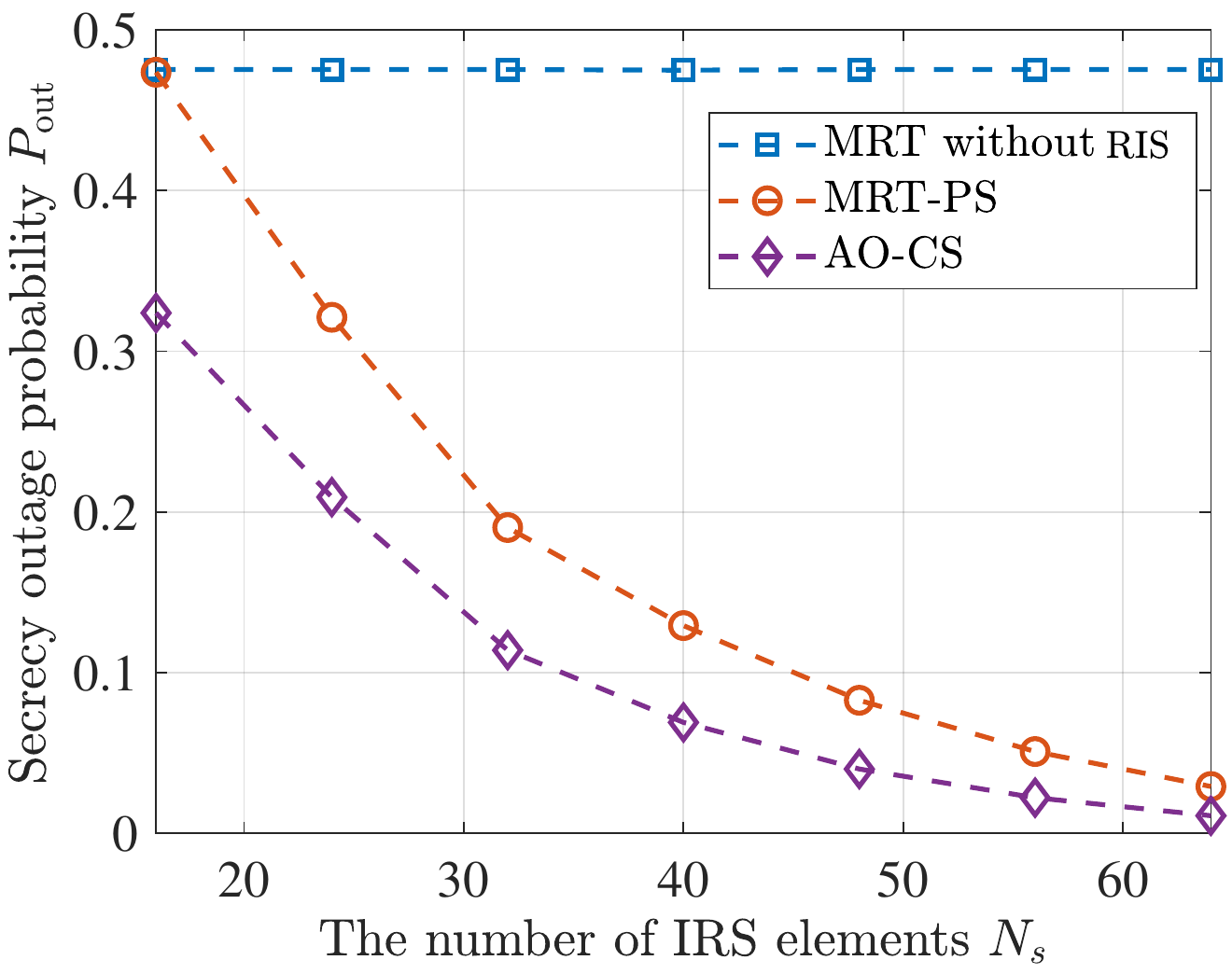}
\caption{$N_t=10$, $N_e=2$, $\alpha=\beta=0.8$, SNR is 9 dB, and $R_s=3$ bit/s/Hz.}\label{simbe}
\end{subfigure}
\caption{Secrecy outage probability in RIS-assisted MISOME scenarios.}
\end{figure*}

Before the comparison simulations, we first verify the convergence performance of the alternating optimization scheme in Fig. \ref{convergence_sim}. As can be seen in this figure, $P_\text{out}$ converges as the number of iterations increases in the AO-CS scheme. The curve shows a good performance of algorithm convergence as the convergence point appears before ten iterations.

The secrecy outage probability in terms of SNR is evaluated in Fig. \ref{simSNR}. Several observations can be made as follows. Firstly, we can find that the secrecy outage probability decreases with an increasing SNR then nearly approaches to a constant in higher SNR regions, and transmission power provides a less gain for RIS-assisted PLS compared to RIS-free cases. Secondly, the RIS-assisted schemes outperform the scheme without RIS as RIS can provide more randomness in wireless channels. Lastly, the proposed AO-CS scheme has a better performance than others.

Next, we want to show the impact of the PLS coding rate on secrecy outage probability in Fig. \ref{simRs}, where an increasing PLS coding rate can enlarge the secrecy outage probability in all schemes. The proposed AO-CS outperforms these schemes, and the advantage is obvious in the higher rate region. The effects of the number of Eve's antennas on secrecy outage probability are examined in Fig. \ref{simNe}, where these curves show an increasing trend in secrecy outage probability when Eve has more antennas, but the lines rise slowly in the AO-CS scheme. It means the eavesdropping antenna effect used to be an important consideration and does much harm on traditional PLS schemes, but is relieved by RIS. Also, we can find that the secrecy outage probability reaches to one in the case of $N_e>N_t$ because Eve has enough antennas to obtain the larger gain than that of Bob, no matter how to change the channels by RIS. The proposed scheme absolutely outperforms others a lot when the number of Eve's antennas is large, e.g., the secrecy outage probability of the proposed scheme is reduced by almost $20\%$ compared to the MRT-PS scheme when $N_e=5$.

Finally, Figs. \ref{simal} and \ref{simbe} illustrate the impacts of the number of Alice's antennas and RIS elements on secrecy outage probability, respectively. From Fig. \ref{simal}, we can observe that a large number of Alice's antennas will decrease secrecy outage probability in all schemes. With phase shifter optimization, the secrecy outage probability decreases quickly as the optimal phase shifter matrix provides a larger gain for the main channel compared to the individual direct channel. A similar trend is seen in Fig. \ref{simbe}, which shows that the secrecy outage probability is reduced with the increasing number of RIS elements. As the cost of RIS elements is less than that of antenna radio frequency modules, a huge number of elements are available in wireless communications. Throughout all simulations, the AO-CS is the best choice for PLS as it significantly reduces the secrecy outage probability with acceptable computational cost.

\begin{figure*}[h!]
\begin{subfigure}[t]{.32\textwidth}
\centering
\includegraphics[width=0.9\textwidth]{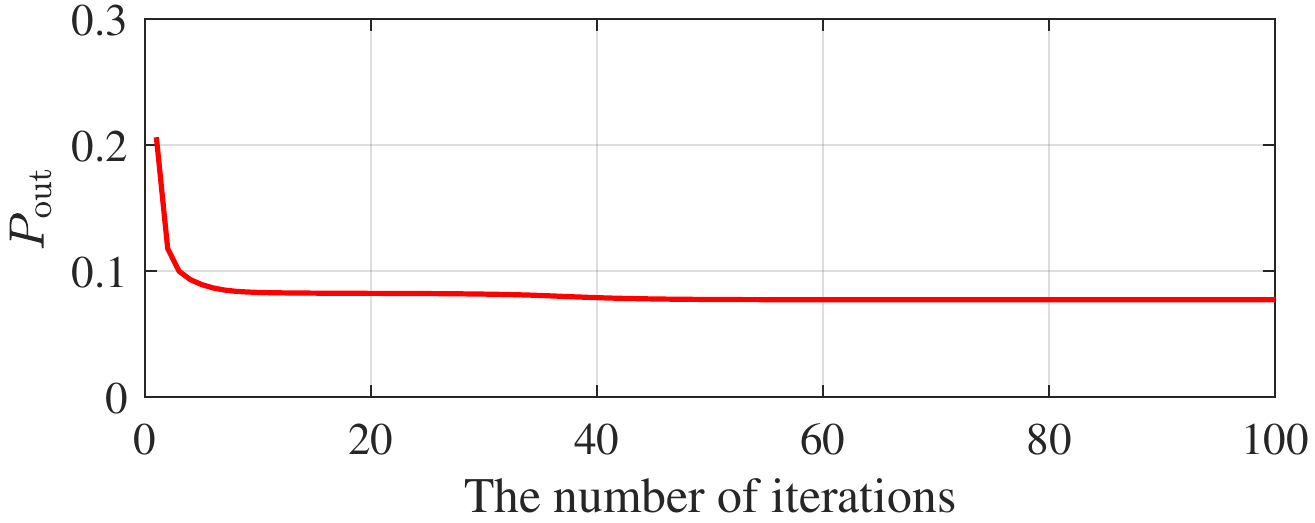}
\caption{AO-Man convergence.}
\end{subfigure}\hfill
\begin{subfigure}[t]{.32\textwidth}
\centering
\includegraphics[width=0.9\textwidth]{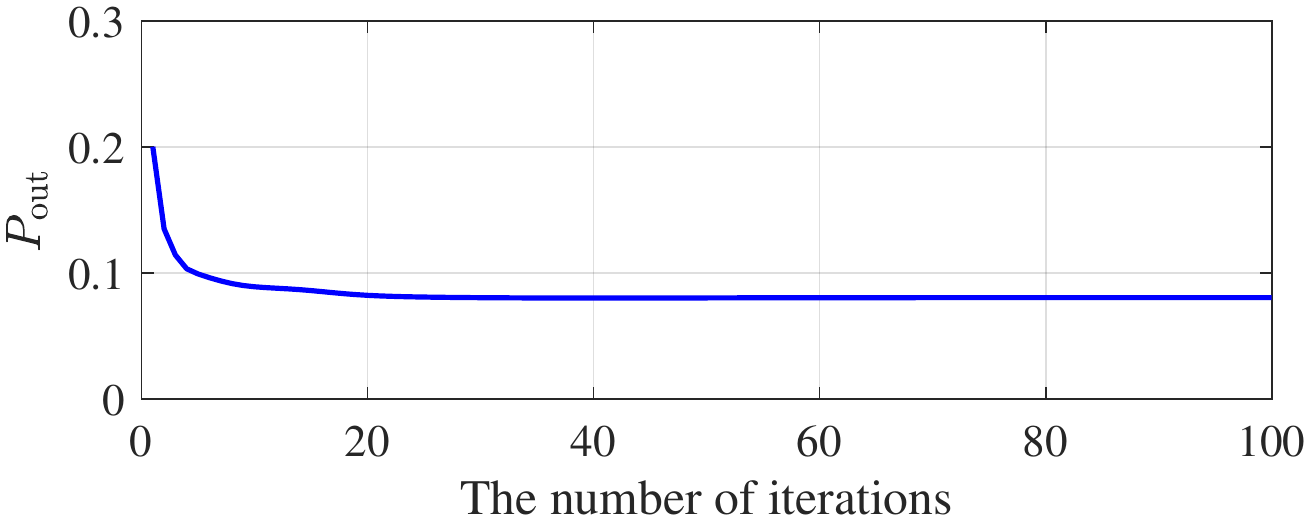}
\caption{AO-SDR convergence.}
\end{subfigure}
\begin{subfigure}[t]{.32\textwidth}
\centering
\includegraphics[width=0.9\textwidth]{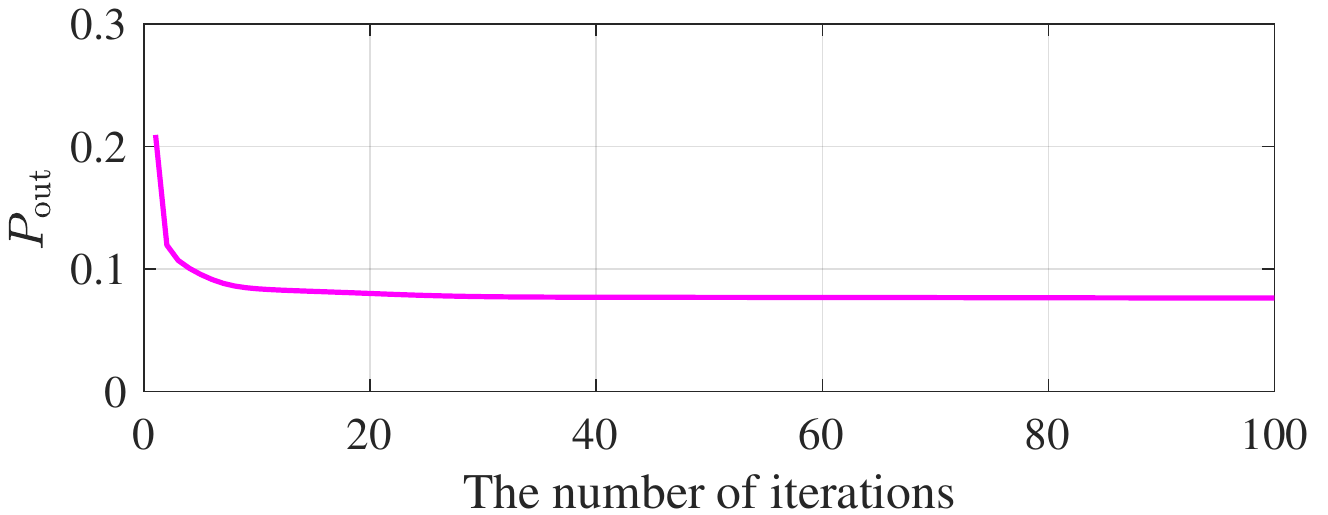}
\caption{AO-MM convergence.}
\end{subfigure}
\caption{Convergence tests in RIS-assisted MIMOME scenarios, where $N_s=32$, $N_e=2$, $N_t=10$, $N_r=3$, $\alpha=\beta=0.8$, SNR is 7 dB, and PLS coding rate is 3 bit/s/Hz.}\label{convergence_sim2}
\end{figure*}

\subsection{Multiple-Antenna Alice and Bob}
In the RIS-assisted MIMOME scenarios, we consider four different schemes, i.e., MRT without RIS, the proposed AO scheme based on SDR (AO-SDR),  manifold optimization (AO-Man), and MM (AO-MM). They are described as follows.

\begin{enumerate}
\item MRT without RIS: Alice performs MRT-based beamforming, i.e., $\mathbf{w}/\sqrt{\rho}=\text{eigvec}_{\lambda_{\max}}(\mathbf{H}_b^{\rm{H}}\mathbf{H}_b)$,  where $\mathbf{H}_b$ represents the channel between Alice and Bob in the scenarios without RIS~\cite{Kang2003}, and phase shift control is not considered. Meanwhile, Bob uses $\mathbf{w}^{\rm{H}}/\sqrt{\rho}$ as the receiving vector. The secrecy outage probability of MRT-based beamforming is calculated as $P_{\text{out,MRT}}(R_s)=\Gamma(N_e,\phi_m)/\Gamma(N_e)$, where $\phi_{m}= \sigma_e^2(2^{C_{m}-R_s}-1)/\rho$ and $C_m=\log_2(1+\rho\alpha^2\lambda_{\max}/\sigma^2)$. The computational complexity of MRT-based beamforming is $O(N_b^3)$.
\item AO-SDR: Alice performs alternating optimization between beamforming and SDR-based phase shifter optimization as described in Section \ref{proposed1}. B. The secrecy outage probability of the proposed AO-SDR scheme can be measured by Eq. (\ref{esop}).  The computational complexity is $O\{\text{iter}_{\max}[\ln(1/\epsilon)N_s^{4.5}+N_t^3+N_s^2N_t]\}$ as discussed in Section \ref{proposed1}. E.
\item AO-Man: Alice performs alternating optimization between beamforming and manifold-based phase shifter optimization as described in Section \ref{proposed1}. C. The secrecy outage probability of the proposed AO-Man scheme can be measured by Eq. (\ref{esop}). The computational complexity is $O[\text{iter}_{\max}(N_s^4+N_t^3+N_s^2N_t)]$ as discussed in Section \ref{proposed1}. E. 
\item AO-MM: Alice performs alternating optimization between beamforming and MM-based phase shifter optimization \cite{HongAN2020}. The secrecy outage probability of the proposed AO-MM scheme\footnote{The detail of  the AO-MM for P2 is shown in: \url{https://github.com/yiliangliu1990/liugit_pub/tree/master/IRS}.} can be measured by Eq. (\ref{esop}). The computational complexity is $O\{\text{iter}_{\max}[N_s^3+N_{\text{iter}}(N_s^2+N_s)]\}$, where $N_{\text{iter}}$ is the number of iterations for the MM algorithm \cite{HongAN2020}. 
\end{enumerate}

Firstly, we also verify the convergence performance of the AO-Man, AO-SDR, and AO-MM schemes in Fig. \ref{convergence_sim2}. It is shown that $P_\text{out}$ converges as the number of iterations increases in all schemes. The curve shows a good performance of algorithm convergence as the convergence point appears before 10 iterations. 

\begin{figure*}[h!]
\begin{subfigure}[t]{.32\textwidth}
\centering
\includegraphics[width=1\textwidth]{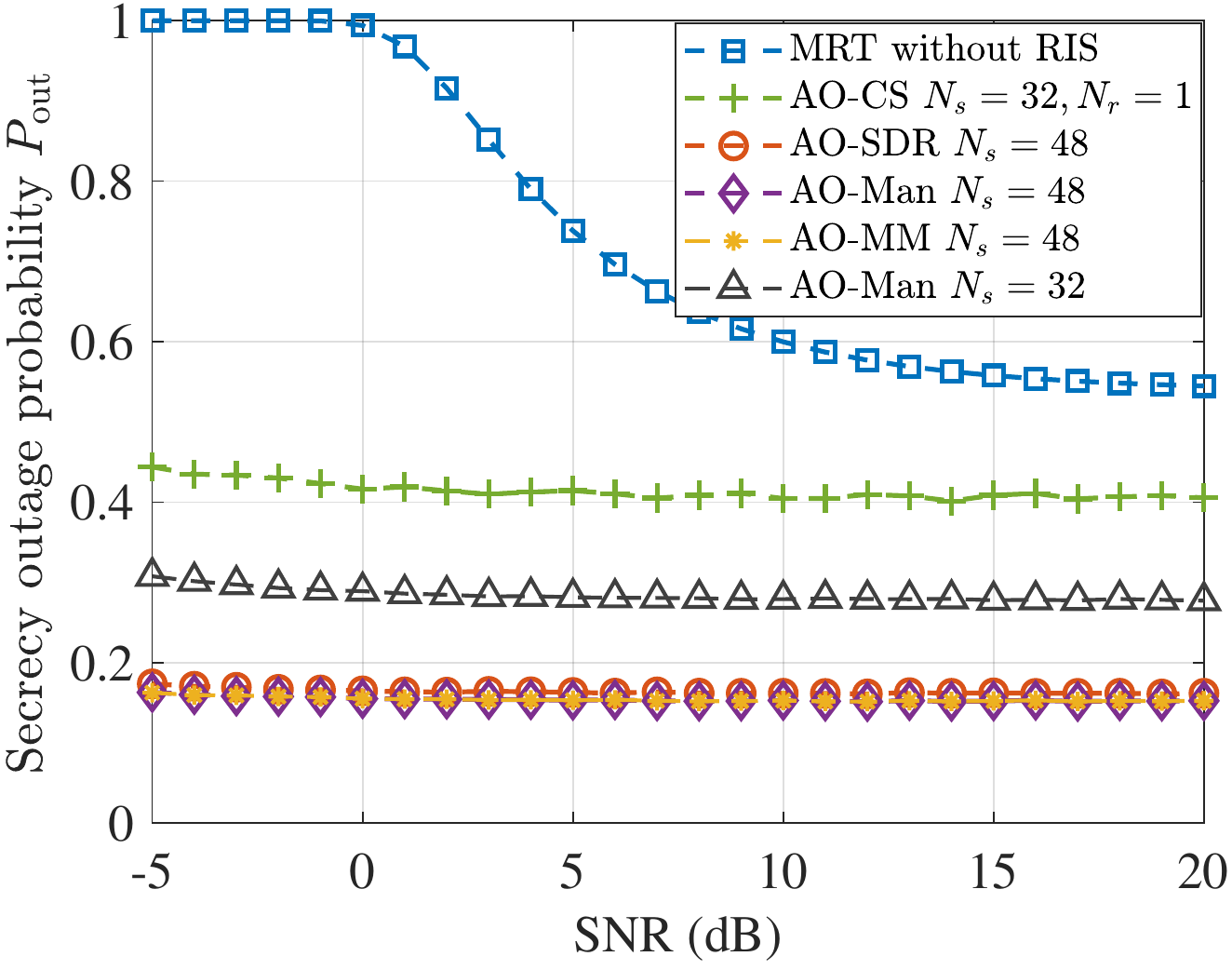}
\caption{$N_t=10$, $N_e=2$, $N_r=3$, $\alpha=\beta=0.8$, and $R_s$ is 4 bit/s/Hz.}\label{mimome1}
\end{subfigure}\hfill
\begin{subfigure}[t]{.32\textwidth}
\centering
\includegraphics[width=1\textwidth]{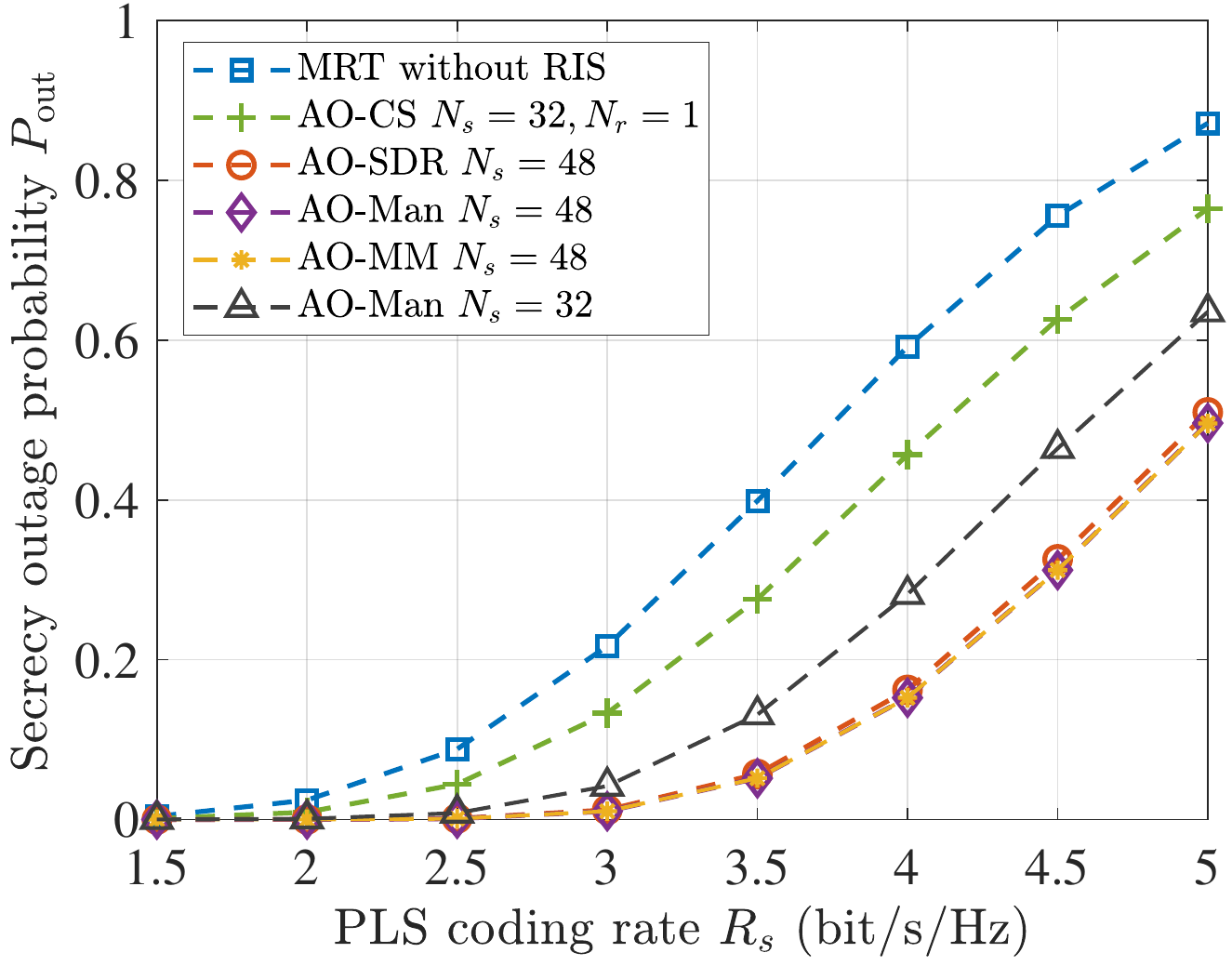}
\caption{$N_t=10$, $N_e=2$, $N_r=3$, $\alpha=\beta=0.8$, and SNR is 9 dB.}\label{mimome2}
\end{subfigure}
\begin{subfigure}[t]{.32\textwidth}
\centering
\includegraphics[width=1\textwidth]{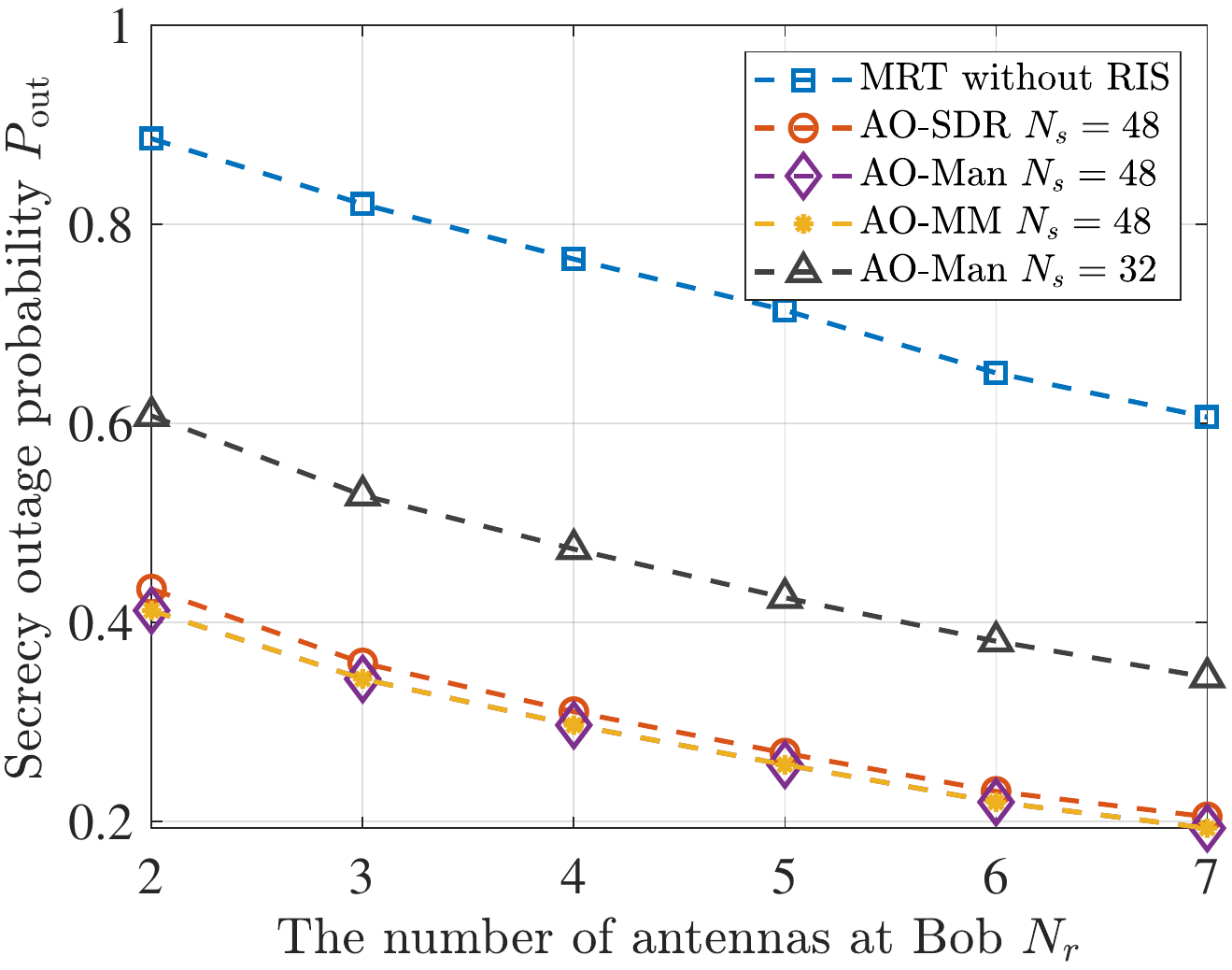}
\caption{$N_t=10$, $N_e=2$, $\alpha=\beta=0.8$, SNR is 9 dB, and $R_s$ is 4 bit/s/Hz.}\label{mimome3}
\end{subfigure}
\caption{Secrecy outage probability in RIS-assisted MIMOME scenarios.}
\end{figure*}

The SNR impact on the secrecy outage probability of RIS-assisted MIMOME models is evaluated in Fig. \ref{mimome1}. Similar to the single-antenna Bob case, the transmission power provides a lower gain in RIS-assisted PLS as we can find that the secrecy outage probability slowly decreases with an increasing SNR. It is shown that the proposed AO-SDR, AO-Man, and AO-MM schemes have a better performance than the MRT without RIS, and the AO-Man and AO-MM are better than the AO-SDR because the maximum eigenvalue method in the AO-SDR can not guarantee to provide the optimal phase shifter matrix, as discussed in Section V. B. Last but not least, by comparing the green line with the black one, we can find the secrecy performance is improved when the multiple antennas are used at Bob. It means that our schemes applied to the RIS-assisted MIMOME case outperform the state-of-the-art researches \cite{Wuirs2019,Hong2019,Dong2020} as they focus on the RIS-assisted MISOME case.

Then, we want to show the effect of the PLS coding rate on secrecy outage probability in Fig. \ref{mimome2}. We find that an increasing PLS coding rate results in the rise of  secrecy outage probability in all schemes, and the proposed schemes outperform much of these schemes. It is also demonstrated that the low computational complexity scheme, i.e., the MRT without RIS, has a similar performance with others when the preset PLS coding rate is small, so it is a possible plan if computational resources are scarce, although it has the worst security performance. Lastly, the impact of the number of Bob's antennas is checked in Fig. \ref{mimome3}. Although we have known the increasing number of Bob's antennas can reduce the secrecy outage probability through the simulations above, Fig. \ref{mimome3} shows a fast linear descent on the secrecy outage probability. It is because MIMO provides a larger diversity gain than the MISO case, and larger randomness that contributes to PLS will be produced among antennas when more antennas are used at Bob.

\section{Conclusions} \label{conclusions}
This paper focuses on the PLS scenario without the instantaneous CSI of eavesdropper in RIS-assisted MIMOME channels. The expression of secrecy outage probability for any beamforming vector and phase shifter matrix is deduced by Gamma distribution. Following that, we design an AO algorithm between the beamforming vector and phase shifter matrix to reduce the secrecy outage probability. Also, the AO algorithm is simplified in the single antenna cases. Simulation results have shown that the proposed scheme can significantly reduce secrecy outage probability compared to state-of-the-art schemes. As one of our future works, we will investigate the multiple-user scenario and multiple data streams in RIS-assisted PLS, which is more suitable for large-scale networks and high data rate requirements.

\section*{Appendix}

\subsection{Proof of Lemma 1}\label{pol1}

Recalling the random variable $x=|\beta\mathbf{a}+\mathbf{Cu}|^2$ as shown in Eq. (\ref{x1}). We begin to calculate the mean and variance of $x\sim X(\beta,m,n,\mathbf{u})$ as follows. At first, the mean of $x$ is expressed as 
\begin{flalign}
\mathbb{E}(x)&=\mathbb{E}(\beta^2|\mathbf{a}|^2)+\mathbb{E}(|\mathbf{Cu}|^2)=m(\beta^2+|\mathbf{u}|^2). \label{meanw2}
\end{flalign}
Then, we will deduce the variance of $x$, i.e., $\text{Var}(x)$, which is given as 
\begin{flalign}\label{var}
\text{Var}(x)=\mathbb{E}(|x|^2)-|\mathbb{E}(x)|^2,
\end{flalign}
where $\mathbb{E}(|x|^2)$ can be transformed as
\begin{flalign}\label{mean1s}
\mathbb{E}(|x|^2)&=\mathbb{E}(\big|| \beta \mathbf{a}+\mathbf{Cu}|^2\big|^2) \notag \\
&=\mathbb{E}\big(\big||\beta\mathbf{a}|^2+|\mathbf{Cu}|^2+ \beta\mathbf{u}^{\rm{H}}\mathbf{C}^{\rm{H}}\mathbf{a}+\beta\mathbf{a}^{\rm{H}}\mathbf{Cu}\big|^2\big) \notag \\ 
&=\mathbb{E}\big(\big||\beta\mathbf{a}|^2\big|^2)+\mathbb{E}(\big||\mathbf{Cu}|^2\big|^2)+\mathbb{E}(|\beta\mathbf{u}^{\rm{H}}\mathbf{C}^{\rm{H}}\mathbf{a}|^2\big) \notag \\
&+\mathbb{E}(|\beta\mathbf{a}^{\rm{H}}\mathbf{Cu}|^2)+2\mathbb{E}(|\beta\mathbf{a}|^2|\mathbf{Cu}|^2),
\end{flalign}
$\mathbb{E}(| \beta \mathbf{u}^{\rm{H}} \mathbf{C}^{\rm{H}}\mathbf{a}|^2 \big)=\mathbb{E}(|\beta \mathbf{a}^{\rm{H}}\mathbf{Cu}|^2)=\beta^2 m|\mathbf{u}|^2$, and $\mathbb{E}(|\beta\mathbf{a}|^2|\mathbf{Cu}|^2)=\beta^2 m^2|\mathbf{u}|^2$. According to the property of noncentral chi-square distribution, the mean and variance  of $|\beta\mathbf{a}|^2$ is $\beta^2 m$ and $\beta^4 m$, respectively. Hence, $\mathbb{E}(\big|| \beta \mathbf{a}|^2\big|^2)$ can be expressed as
\begin{flalign}
\mathbb{E}(\big||\beta\mathbf{a}|^2\big|^2)&= \text{Var}(|\beta\mathbf{a}|^2)+[\mathbb{E}(|\beta\mathbf{a}|^2)]^2=\beta^4(m+m^2).
\end{flalign}
We introduce an auxiliary random variable $\mathbf{z}_1=\mathbf{Cu}/|\mathbf{u}|$ such that $\mathbf{z}_1\sim\mathcal{CN}_{m,1}(\mathbf{0},\mathbf{I}_{m})$. We change the form of $\mathbb{E}(\big||\mathbf{Cu}|^2\big|^2)$ as
\begin{flalign}
\mathbb{E}(\big||\mathbf{Cu}|^2\big|^2)&=\mathbb{E}(|\mathbf{u}|^4|\mathbf{z}_1|^4) =(m^2+m)|\mathbf{u}|^4.
\end{flalign}
Following that, we have
\begin{flalign}
\mathbb{E}(|x|^2)=(\beta^4+|\mathbf{u}|^4+2\beta^2|\mathbf{u}|^2)(m^2+m).
\end{flalign}
Then, according to Eq. (\ref{var}), $\text{Var}(x)$ can be expressed as
\begin{flalign}
\text{Var}(x)=(\beta^4+|\mathbf{u}|^4+2 \beta^2|\mathbf{u}|^2)m.
\end{flalign}
Hence, the shape and scale of the Gamma distribution can be expressed as \cite{weisstein2002crc}
\begin{flalign}\label{gammapara}
&k=\frac{[\mathbb{E}(x)]^2}{\text{Var}(x)}=m, \quad w=\frac{\text{Var}(x)}{\mathbb{E}(x)}=\beta^2 +|\mathbf{u}|^2.
\end{flalign}
At last, on the basis of the definition of the Gamma distribution \cite{weisstein2002crc}, we get the CDF of $X$ as follows,
\begin{flalign}
F_{X}(x)&=1-\frac{1}{\Gamma(k)}\Gamma(k,\frac{x}{w}) \notag \\
&=1-\frac{1}{\Gamma(m)}\Gamma(m,\frac{x}{\beta^2+|\mathbf{u}|^2}),
\end{flalign}
where $\Gamma(x)$ is the Gamma function of variable $x$, and $\Gamma(\epsilon,\eta)$ is the upper incomplete Gamma function defined in Eq. (\ref{ligamma}). The proof is completed. \hfill $\IEEEQEDclosed$

\bibliographystyle{IEEEtran}
\bibliography{references}
\end{document}